\documentclass[aps,twocolumn,showpacs,amsmath,amssymb,pra,superscriptaddress,floatfix,longbibliography]{revtex4-2}

\usepackage{braket}
\usepackage{amsmath}
\usepackage{amssymb}
\usepackage{mathtools}
\usepackage{graphicx}
\usepackage{dcolumn}
\usepackage{bm}
\usepackage{multirow}
\usepackage{appendix}
\usepackage{leftidx}
\usepackage{color}
\usepackage{float}
\usepackage{qcircuit}
\usepackage{ifthen}
\usepackage{tabularx}
\usepackage{amsfonts}
\usepackage[german,english]{babel}
\usepackage{soul} 
\usepackage{comment}
\usepackage{afterpage}

\newcommand {\rsub}[1]{\textcolor{black}{#1}}

\newcommand \be{\begin{equation}}
\newcommand \ee{\end{equation}}
\newcommand \bea{\begin{eqnarray}}
\newcommand \eea{\end{eqnarray}}
\newcommand \bse{\begin{subequations}}
\newcommand \ese{\end{subequations}}

\usepackage{xcolor}
\newboolean{ShowComments}
\setboolean{ShowComments}{false}  

\definecolor{mscolor}{rgb}{0,0.5,0.5}

\definecolor{xjcolor}{rgb}{0.5,0,0.5}

\newcommand{\InfleqtionM}{Infleqtion, Inc., Madison, WI, 53703, USA}
\newcommand{\UWM}{Department of Physics, University of Wisconsin-Madison, 1150 University Avenue, Madison, WI, 53706, USA}

\newcommand{\Forth}{
Institute of Electronic Structure and Laser, FORTH, GR-70013 Heraklion, Crete, Greece}

\begin{document}

\title{
Fast measurements and multiqubit gates in dual species atomic arrays}
\author{D. Petrosyan}
\affiliation{\Forth}
\author{S. Norrell}
\affiliation{\UWM}
\author{C. Poole}
\affiliation{\UWM}
\author{M. Saffman}
\affiliation{\UWM}
\affiliation{\InfleqtionM}

\date{\today}

\begin{abstract}
We propose and analyze an approach for fast syndrome measurements in an array of rubidium and cesium atomic qubits. 
The scheme works by implementing an inter-species  $\textsf{CNOT}_k$ gate, entangling one cesium ancilla qubit with $k\geq 1$ rubidium qubits which are then used for state measurement. 
Utilizing Rydberg states with different inter- and intra-species interaction strengths, the proposal provides  a syndrome measurement fidelity of $\mathcal{F}>0.9999$ in less than 5 $\mu$s of integration time.
\end{abstract}
 
\maketitle

\section{Introduction}

Fast and accurate mid-circuit syndrome measurements are an essential requirement for quantum error correction.
Recent experiments have demonstrated several complementary approaches to mid-circuit qubit measurements in neutral atom arrays. These include atom transport to prevent scattered light from disturbing proximal data qubits \cite{Deist2022,Bluvstein2024}, shelving data qubits  in states that are dark to readout light \cite{Graham2023b,SMa2023,Norcia2023,Lis2023}, and a two-species approach that spectrally isolates measurements of atoms used as ancilla qubits from atoms used as data qubits \cite{Singh2023}. 
Although quantum gate operations based on Rydberg interactions have fast sub-$\mu\rm s$ timescales, qubit state measurements in free space are typically orders of magnitude slower with few ms timescales. In order to reach fast cycle times for quantum error correction of logical qubits, it is important to decrease state measurement times as much as possible, and to do so without causing atom loss or crosstalk affecting other qubits. 
 
Measurement of the atomic state can be realized using several approaches including imaging of scattered light, and detection of absorption or optical phase shifts imparted by an atom. Near-resonant scattering  is the most widely used approach whereby atoms in a bright state $\ket{1}$ scatter many photons that are detected, while atoms in a far detuned dark state $\ket{0}$ have orders of magnitude lower scattering rates. The time needed to measure the quantum state with a desired fidelity depends on the atomic scattering rate, the photon detection efficiency, and the background count rate which includes undesired optical scattering from surfaces and detector dark counts \cite{Saffman2005a}.  
In addition to having fast and high fidelity measurements, it is important that these measurements are lossless to avoid the need to replace ancilla atoms after a measurement. 
The effective measurement rate must also account for the time needed to reset the ancilla and measuring atoms for  the next measurement cycle. 

One way to reduce the measurement time is to map the quantum state of an ancilla onto more than one auxiliary measuring atom to obtain a stronger optical signal. 
Such an idea was introduced in \cite{Saffman2005b} and an 
approach using electromagnetically induced transparency (EIT) was demonstrated in \cite{WXu2021}. Both of these previous methods require 100-1000 auxiliary atoms for optimal performance. 
We suggest here an approach that can reach measurement times under $10~\mu\rm s$ and measurement fidelity better than 0.9999, with only a few auxiliary atoms. 

\begin{figure}[b]
   \includegraphics[width=0.9 \columnwidth]{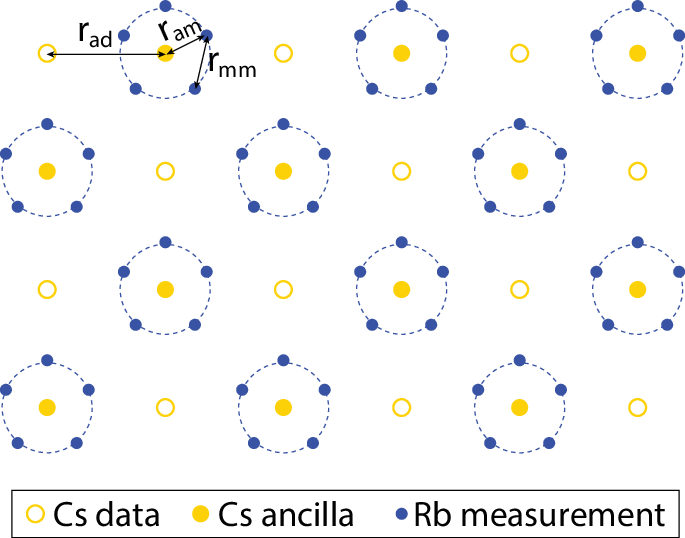}
   \caption{Surface code layout with Cs data and ancilla qubits and Rb measuring atoms. The relative distances are $r_{\rm da}$ - data to ancilla qubits, $r_{\rm am}$ - ancilla to measurement qubits, and $r_{\rm mm}$ - measurement to measurement qubits. }
    \label{fig.surfacecode}
\end{figure}

The proposed arrangement of atoms for surface code implementation is shown in Fig.~\ref{fig.surfacecode}. Data and ancilla Cs atom qubits are placed on a two-dimensional (2D) grid \cite{Fowler2012}. 
Rydberg quantum gates between the nearest neighbor data and ancilla qubits are used to map $\sf X$ and $\sf Z$ parity syndromes onto the ancillae. The state of each ancilla atom is then mapped onto $k$ Rb measuring atoms, which are subsequently interrogated using fluorescence imaging.
The desired mapping between ancillae (a) and measuring (m) qubits is 
\bse\bea
\ket{0}_{\rm a}\ket{ 0}_{\rm m}^{\otimes k}&\rightarrow& \ket{0}_{\rm a}\ket{ 0}_{\rm m}^{\otimes k},\\
\ket{1}_{\rm a}\ket{0}_{\rm m}^{\otimes k}&\rightarrow& \ket{1}_{\rm a}\ket{1}_{\rm m}^{\otimes k}.
\eea
\label{eq.mapping}
\ese
The necessary arrangement of optical traps can be generated with a spatial light modulator \cite{Barredo2016,Kim2016,DKim2019}, or a pre-fabricated mask \cite{Huft2022}. Since the measuring atoms are a different species than the data atoms, and therefore have  very different resonant optical wavelengths (852 nm for Cs and 780 nm for Rb), there is essentially no dephasing of the Cs data qubits during measurement \cite{Beterov2015,Singh2023}.  

The level scheme for the state mapping between the Cs ancilla and Rb measuring atoms is shown in Fig.~\ref{fig:scheme}. \rsub{
The mapping is based on a modified form of  Rydberg controlled state transfer \cite{Saffman2009b,Jau2016}.}
The protocol operates as follows. 
Rb atoms are prepared in state $\ket{0}_{\rm m}$ which is dark to the interrogating laser. The Cs ancilla is either in state $\ket{0}_{\rm a}$ and remains there, or in state $\ket{1}_{\rm a}$ and is transferred to the Rydberg state $\ket{r}_{\rm a}$.
A microwave $\pi$-pulse resonant with the transition $\ket{0}_{\rm m} \to \ket{1}_{\rm m}$ is applied to the Rb atoms. State $\ket{1}_{\rm m}$ is also coupled to a Rydberg state $\ket{r}_{\rm m}$ by a non-resonant laser leading to a dynamic Stark shift of $\ket{1}_{\rm m}$ that depends on whether Cs atom is in non-interacting state $\ket{0}_{\rm a}$ or in the strongly-interacting Rydberg state $\ket{r}_{\rm a}$. In the former case, the Stark shift is large making the microwave pulse non-resonant and thus suppressing the transfer $\ket{0}_{\rm m} \to \ket{1}_{\rm m}$. In the latter case, the Rydberg level $\ket{r}_{\rm m}$ is largely shifted by the strong interaction with the Cs atom, which in turn reduces the Stark shift of $\ket{1}_{\rm m}$ making the transition $\ket{0}_{\rm m} \to \ket{1}_{\rm m}$ resonant with the microwave pulse, transferring the Rb atoms to $\ket{1}_{\rm m}$. 
This implements the mapping described in Eqs.~(\ref{eq.mapping}). 

After the state mapping is performed, the Rb measuring atoms are subjected to an interrogating laser acting on the cycling transition $\ket{1}_\mathrm{m} \to \ket{f}_\mathrm{m}$ and their state is recorded. The measuring atoms optically pumped to $\ket{1}_{\rm m}$ can then be rapidly returned to the initial state $\ket{0}_{\rm m}$ by another microwave $\pi$-pulse. The Cs ancillae also need to be reset. This might appear to be problematic since scattering of repump photons from the Cs ancillae could cause dephasing of nearby Cs data qubits; however, the measurement results tell us which state the Cs ancillae are in and local $\pi$ pulses on the qubit transition can be used to reset the ancillae as needed, so optical pumping of the Cs atoms is not required.   

The mapping of Eqs.~(\ref{eq.mapping}) is also equivalent to the $\textsf{CNOT}_k$ gate with one Cs control qubit and $k$ Rb target qubits. Implementing this gate with high fidelity is, however, more demanding experimentally but still possible with our scheme, as discussed below.

The paper is organized as follows. 
In Sec.~\ref{sec.dynamics} we provide a quantitative analysis of the performance of the state mapping, followed by an estimate of the achievable measurement and quantum error correction cycle time, including ancilla reset,  in Sec.~\ref{sec.time}. The  multiqubit $\textsf{CNOT}_k$ gate is analyzed in Sec.~\ref{sec.gate}. We conclude in Sec.~\ref{sec.outlook} with a summary and outlook. 
Rydberg parameters used in the analysis are documented in the Appendix.

\section{Dynamics of the measuring atoms}
\label{sec.dynamics}

\begin{figure}[!t]
\includegraphics[width=0.9\columnwidth]{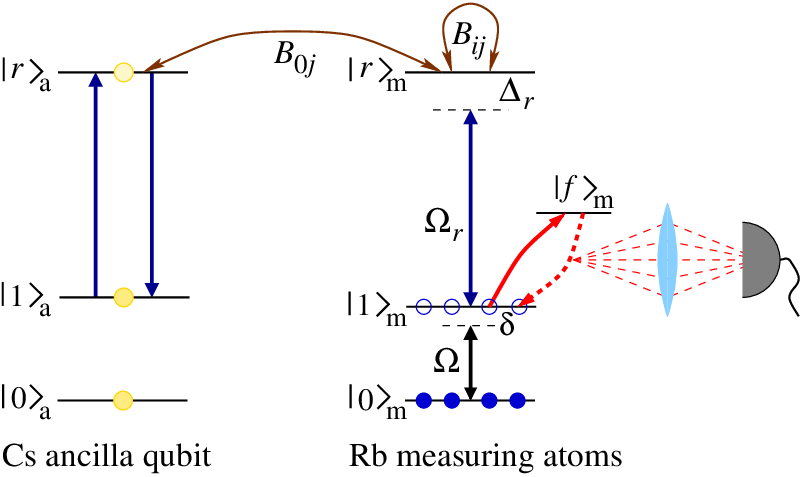}
\caption{Level scheme and couplings of the Cs ancilla qubit and Rb measuring atoms. 
Conditional upon the state of the ancilla qubit $\ket{0,1}_{\mathrm{a}}$, the Rb atoms are transferred to state $\ket{1}_{\mathrm{m}}$ and then interrogated by a laser on the cycling transition to state $\ket{f}_\mathrm{m}$ providing a fluorescence signal for measurement.}
\label{fig:scheme}
\end{figure}

We proceed to describe the system in more detail to quantify the measurement time and fidelity. 
The qubit states of the Cs ancilla atom are $\ket{0}_\mathrm{a} \equiv \ket{6s_{1/2}, f=3,m_f=0}$ and $\ket{1}_\mathrm{a} \equiv \ket{6s_{1/2},f=4,m_f=0}$. 
To measure the state of the ancilla qubit, a two-photon resonant laser pulse of area $\pi$ can transfer the population of state $\ket{1}_\mathrm{a}$ to the Rydberg state $\ket{r}_\mathrm{a} \equiv \ket{48s_{1/2},m_j=1/2}$ that interacts strongly with the corresponding Rydberg state of the measuring Rb atoms, as detailed below and illustrated in Fig.~\ref{fig:scheme}. 

We denote by $\ket{0}_{\rm m} \equiv \ket{5s_{1/2}, f=1,m_f=1}$ and $\ket{1}_{\rm m} \equiv \ket{5s_{1/2},f=2,m_f=2}$ the ground-state hyperfine sublevels of $^{87}$Rb atoms, and the Rydberg state  $\ket{r}_{\rm m} \equiv \ket{46s_{1/2},m_j=1/2}$. 
A near-resonant microwave field acts on the transition $\ket{0}_{\rm m} \rightarrow  \ket{1}_{\rm m}$ with  Rabi frequency $\Omega$ and detuning $\delta$, while state $\ket{1}_{\rm m}$ is non-resonantly coupled to, or dressed with, the Rydberg state $\ket{r}_{\rm m}$ by a pair of lasers with the two-photon Rabi frequency $\Omega_r$ and detuning $\Delta_r \gg \Omega_r$. The corresponding  Hamiltonian ($\hbar=1$), in the frame rotating with the microwave and laser frequencies, is given by 
\begin{eqnarray}
H_{\mathrm{m}} &=&  \sum_j \Big[ \delta \hat\sigma_{11}^{(j)} + \big( \delta+\Delta_r - \tfrac{i}{2} \gamma_r \big) \hat\sigma_{rr}^{(j)} 
\nonumber \\
&&\qquad -  \tfrac{1}{2}(\Omega \hat\sigma_{10}^{(j)} + \Omega_{ r} \hat\sigma_{r1}^{(j)} + \mathrm{H.c.}) \Big]
\nonumber \\
&+&  \sum_{ij} B_{ij} \hat\sigma_{rr}^{(i)} \sigma_{rr}^{(j)} + \sum_{j} B_{0j} \hat\sigma_{rr}^{(0)} \sigma_{rr}^{(j)}  , 
\label{eq:ham}
\end{eqnarray}
where $\hat\sigma_{\mu \nu}^{{(j)}} \equiv \ket{\mu}_j \bra{\nu}$ are the transition ($\mu\neq\nu$) or projection ($\mu = \nu$) operators for the Rb atoms $j=1,2,\ldots$, 
$\gamma_r$ is the relaxation rate of the Rydberg state (assumed to decay mostly to states other than $\ket{0,1}_\mathrm{m}$), 
$B_{ij}$ is the interaction strength between the Rb atoms $i,j$ in the Rydberg state, and $B_{0j}$ is the interaction strength between the Cs atom (index 0) and the Rb atom $j$. 

Assuming large detuning $\Delta_r > \Omega_r$ and neglecting for now the Rydberg state decay and interactions between the Rb atoms, we can adiabatically eliminate their Rydberg state $\ket{r}_{\rm m}$ obtaining an effective two-level atom Hamiltonian
\begin{equation}
H_{\mathrm{eff}} =  \sum_j \left[ (\delta - \hat\Delta_1^{(j)}) \sigma_{11}^{(j)} - \tfrac{1}{2} (\Omega \sigma_{10}^{(j)} + \mathrm{H.c.}) \right] , \label{eq:hameff}
\end{equation}
with an operator-valued detuning
\begin{equation}
\hat\Delta_1^{(j)} \equiv \frac{|\Omega_r|^2}{4 \left(\Delta_r+B_{0j} \hat\sigma_{rr}^{(0)} \right)} ,
\end{equation}
corresponding to the ac Stark shift of level $\ket{1}_\mathrm{m}$. 
This level shift is sizable, $\hat\Delta_1 \to \frac{|\Omega_r|^2}{4 \Delta_r}$, if the Cs atom was initially in state $\ket{0}_{\rm a}$ and remained there ($\hat\sigma_{rr}^{(0)} \to 0$). 
In contrast, if the Cs atom was initially in state $\ket{1}_{\rm a}$, and was excited to the Rydberg state $\ket{r}_{\rm a}$ ($\hat\sigma_{rr}^{(0)} \to 1$), the level shift is much smaller, $\hat\Delta_1 \to \frac{|\Omega_r|^2}{4 (\Delta_r+B_{0j})}$, for sufficiently large interaction $B_{0j} \gg \Delta_r$.

We then apply a resonant ($\delta\simeq 0$) microwave pulse $\Omega(t)$ of area $\int \Omega(t) dt = \pi$ and peak amplitude $\Omega_\mathrm{max} < \frac{|\Omega_r|^2}{4 \Delta_r}$. If the Cs atom is in the Rydberg state, $\Delta_1 \ll \Omega_\mathrm{max}$, the microwave field is resonant with the transtion $\ket{0}_\mathrm{m} \to \ket{1}_\mathrm{m}$ of the Rb atoms, leading to the population transfer to $\ket{1}_\mathrm{m}$. 
But if the Cs atom is not in the Rydberg state, $\Delta_1 > \Omega_\mathrm{max}$, the microwave field is non-resonant and the transition $\ket{0}_\mathrm{m} \to \ket{1}_\mathrm{m}$ is suppressed, leaving the Rb atoms in state $\ket{0}_\mathrm{m}$. 
We note that since the Rydberg state population $\braket{\sigma_{rr}}$ of the Rb atoms is small at all times, $\max \braket{\sigma_{rr}} \leq \frac{|\Omega_r|^2}{4 \Delta_r^2} \ll 1$, their decay and mutual interactions are suppressed and can therefore be neglected.
Finally, the Rb atoms are interrogated by fluorescence detection on the cycling transition $\ket{1}_\mathrm{m} \to \ket{f}_\mathrm{m}$, where $\ket{f}_\mathrm{m} \equiv \ket{5p_{3/2},f=3,m_f=3}$ \cite{Kwon2017}. 

\begin{figure}[!t]
\includegraphics[width=1.0\columnwidth]{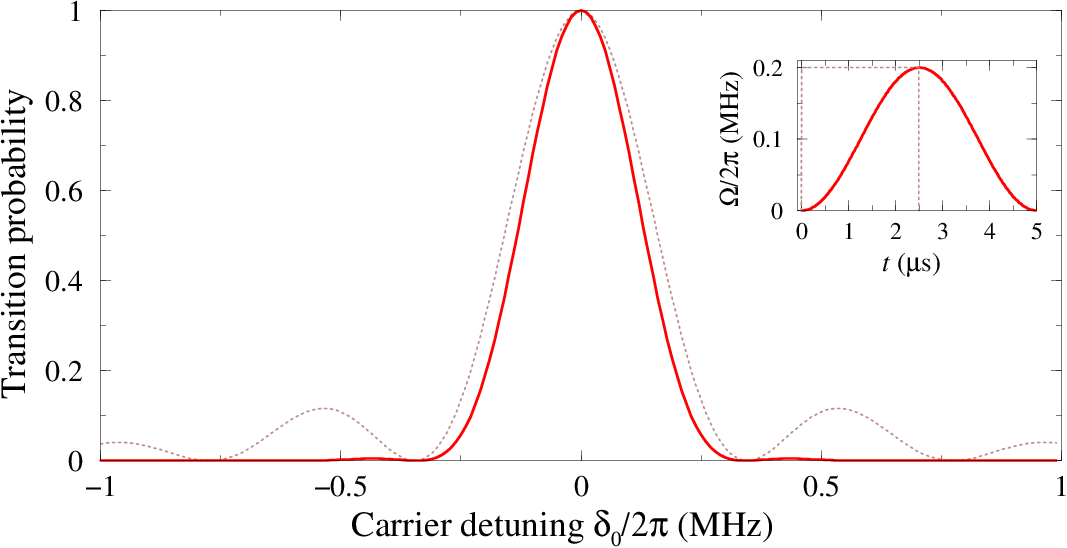}
\caption{Transition probability $\braket{\sigma_{11}(t=\tau)}$ from state $\ket{0}$ to state $\ket{1}$ for a two-level atom driven by 
a square $\pi$-pulse $\Omega(t)=\Omega_\mathrm{max} \Theta(t) \Theta(\tau-t)$ of duration $\tau = \pi/\Omega_\mathrm{max}$ (dotted brown line), and 
a smooth $\pi$-pulse $\Omega(t)=\Omega_\mathrm{max} \sin^2(\pi t/\tau)$ of duration $\tau = 2\pi/\Omega_\mathrm{max}$ (solid red line), vs the detuning $\delta=\delta_0$. 
Inset shows the pulse amplitudes $\Omega(t)$ with $\Omega_\mathrm{max}=2\pi\times 0.2\;$MHz. }
\label{fig:tla}
\end{figure}

We have verified these conclusions by exact numerical simulations for several Rb atoms governed by the Hamiltonian (\ref{eq:ham}).
Before presenting these results, we may  develop some intuition for the required conditions by first examining the transition probability $\braket{\sigma_{11}}$ between states $\ket{0}$ and $\ket{1}$ of a single two-level atom upon applying microwave pulses of different envelopes $\Omega(t)$ and fixed detunings $\delta = \delta_0$, see Fig.~\ref{fig:tla}.
For any pulse of area $\int \Omega(t) dt = \pi$, the transition probability is peaked $\braket{\sigma_{11}} = 1$ around the resonance $\delta_0 = 0$, but away from resonance, $\braket{\sigma_{11}}$ is sensitive to the pulse envelope.  
Thus, for a square pulse, the excitation spectrum 
\[
\braket{\sigma_{11}} = \frac{1}{1 + (\delta/\Omega_\mathrm{max})^2} \sin^2 \left( \tfrac{\pi}{2} \sqrt{1 + (\delta/\Omega_\mathrm{max})^2 }\right)
\]
has long and oscillating tails.
But for a sufficiently smooth $\pi$-pulse, such as, e.g., 
$\Omega(t)=\Omega_\mathrm{max} \sin^2(\Omega_\mathrm{max} t/2)$
($0 \leq t  \leq 2\pi /\Omega_\mathrm{max}$), the tails are highly suppressed, since for a non-resonant pulse the atom initially in state $\ket{0}$ adiabatically follows the instantaneous eigenstate and returns to the same state $\ket{0}$ at the end of the pulse \cite{Petrosyan2017}. 
We will therefore use smooth microwave pulses to conditionally transfer the Rb atoms between the states $\ket{0}_\mathrm{m}$ and $\ket{1}_\mathrm{m}$.

We now consider the complete system consisting of an ancilla Cs atom surrounded by five measuring Rb atoms.
The ancilla atom is either in state $\ket{0}_\mathrm{a}$ or state $\ket{1}_\mathrm{a}$ transferred to the Rydberg state $\ket{r}_\mathrm{a}$ by an appropriate laser pulse(s). 
The measuring atoms are positioned radially around the ancilla atom at distance $r_{\mathrm{am}} \simeq 2.5\;\mu$m. Then the total (dipole-dipole and van der Waals) interaction strength between the Cs and Rb atoms in the Rydberg states $\ket{r}_\mathrm{a}$ and $\ket{r}_{\rm m}$  is $B_{0j} \simeq 2\pi \times 156\;$MHz, while the interaction strength between the neighboring Rb atoms $i,j$ separated by $r_{\mathrm{mm}} \simeq 2.9\;\mu$m is $B_{ij} \simeq 2\pi \times 9\;$MHz. 
State $\ket{1}_{\rm m}$ of the Rb atom is coupled to the Rydberg state $\ket{r}_{\rm m}$ by  lasers with the effective Rabi frequency $\Omega_r/2\pi = 6\:$MHz and large detuning $\Delta_r/2\pi = 15~$MHz, while the transition $\ket{0}_{\rm m}\to \ket{1}_{\rm m}$ is driven by a resonant microwave $\pi$-pulse with a smooth envelope $\Omega(t)=\Omega_\mathrm{max} \sin^2(\pi t/\tau)$ with the peak amplitude $\Omega_{\mathrm{max}}/2\pi = 0.2\;$MHz and duration $\tau = 2\pi/\Omega_\mathrm{max}$, as in Fig.~\ref{fig:tla}.  

\begin{figure}[!t]
\includegraphics[width=1.0\columnwidth]{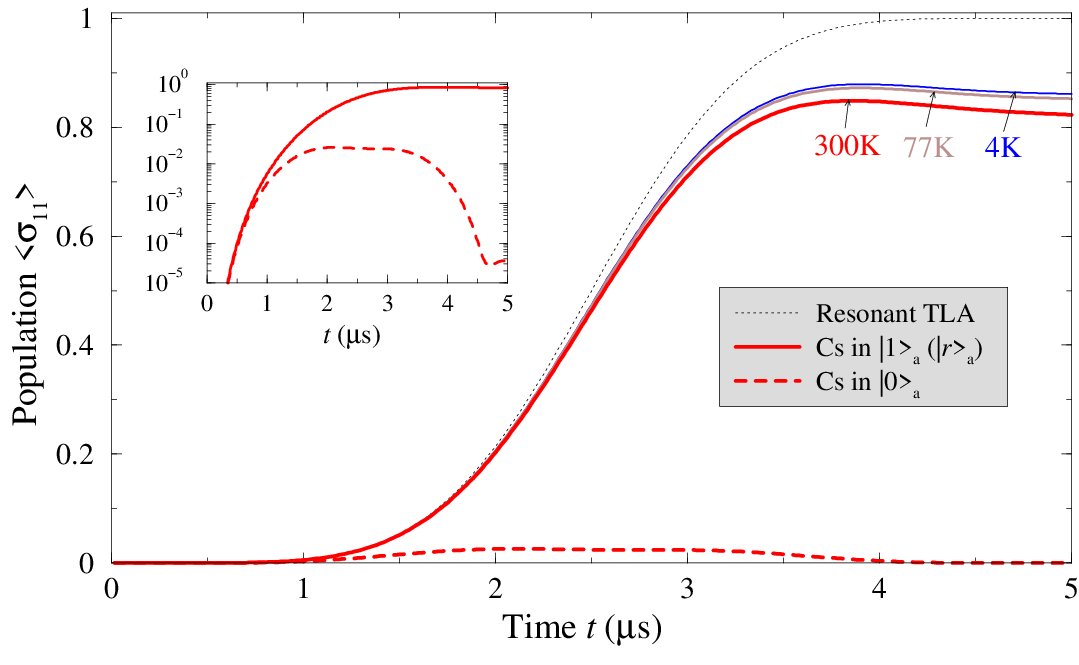}
\caption{Dynamics of populations $\braket{\sigma_{11}^{(j)}}$ of state $\ket{1}_{\rm m}$ of each of the five measuring Rb atoms, initially in state $\ket{0}_{\rm m}$, upon applying a resonant microwave $\pi$-pulse $\Omega(t)=\Omega_\mathrm{max} \sin^2(\pi t/\tau)$ with $\Omega_\mathrm{max}=2\pi\times 0.2\;$MHz. 
State $\ket{1}_{\rm m}$ is dressed with the Rydberg state $\ket{r}_{\rm m}$ by a laser with the Rabi frequency $\Omega_r/2\pi = 6\;$MHz and detuning $\Delta_r/2\pi = 15\:$MHz. 
For the Cs atom in Rydberg state $\ket{r}_\mathrm{a}$, the Rb atoms undergo population transfer to state $\ket{1}_{\rm m}$ (solid lines); while for the Cs atom in state $\ket{0}_\mathrm{a}$, the transfer is suppressed (dashed lines).
In the simulations, we take the Rydberg state lifetimes $1/\gamma_{r_{\rm m}}= 50, 80, 100\:\mu \mathrm{s}$ for Rb and $1/\gamma_{r_{\rm a}}=55, 90, 110\:\mu \mathrm{s}$ for Cs, assuming the temperatures $T=300,77,4\:$K, respectivelly.
For comparison, we also show on the main panel the transfer dynamics for a resonant two-level atom (black dotted line).
Inset shows the populations $\braket{\sigma_{11}^{(j)}}$ of the Rb atoms (for $T=300\:$K) on a log scale. }
\label{fig:Rb01trdyn}
\end{figure}

In Fig.~\ref{fig:Rb01trdyn} we show the dynamics of population transfer between the states $\ket{0}_{\rm m}$ and $\ket{1}_{\rm m}$ of the measuring Rb atoms, as obtained from the numerical solution of the Schr\"odinger equation \rsub{$i\partial_t \ket{\Psi} = H \ket{\Psi}$ for the wavevector $\ket{\Psi}$ of the system governed by the non-Hermitian Hamiltonian (\ref{eq:ham}), assuming that decay of the Rydberg states of Cs and Rb atoms leads to the loss of population to states outside the computational basis.}
The numerical results show that when the Cs atom is in state $\ket{0}_{\rm a}$ the transfer of the measurement atoms is suppressed to less than $10^{-4}$. When the Cs atom is in state $\ket{1}_{\rm a}$ the measurement atom transfer is resonant and succeeds with probability greater than 80\%. 
The combination of the very high fidelity of suppression (ancilla in $\ket{0}_{\rm a}$) and high probability of multi-atom transfer (ancilla in $\ket{1}_{\rm a}$) leads to low measurement error with a short integration time, as detailed in Sec.~\ref{sec.time} below. 

We note again that due to the small population of the Rydberg states of the Rb atoms, $\max \braket{\sigma_{rr}} \ll 1$, their decay (with probability $\gamma_{r} \tau \frac{|\Omega_r|^2}{4\Delta_r^2}$) and interactions do not play a noticeable role in the dynamics of the system. 
But during the transfer, the decay of the Rydberg state of the Cs atom (with probability $\gamma_{r_{\rm a}} \tau$) significantly reduces the transition probability $\braket{\sigma_{11}}$ to state $\ket{1}_\mathrm{m}$ of the measuring Rb atoms. 
This problem can be partially addressed by working at cryogenic conditions to increase the Rydberg state lifetime $1/\gamma_{r_{\rm a}}$ of the Cs atoms.

Another potential concern is the possible ionization of Rydberg states by the 6.8 GHz microwave field used to drive the $\ket{0}_{\rm m}\rightarrow \ket{1}_{\rm m}$ transition. Ionization of non-hydrogenic Rydberg states can occur at a relatively low threshold field of  $ E_{\rm at}/(3 n^5)$ with $E_{\rm at}=5.1\times 10^{11}~\rm V/m$ \cite{Pillet1984}. 
However, for a circularly polarized microwave field, as is used to drive the $\ket{0}_{\rm m}\rightarrow \ket{1}_{\rm m}$ transition, the ionization threshold is $E_{\rm ionize}=E_{\rm at}/(16 n^4)$ \cite{PFu1990}. At a microwave Rabi frequency of 
$\Omega_\mathrm{max}=2\pi\times 0.2\;$MHz the field strength is $E_{\rm max}=3500\:$V/m which is much less than the $n=48$ ionization field of $E_{\rm ionize}=6050\:$V/m. This field strength corresponds to about 1.6 W of microwave power in an area of $1~\rm cm^2$ and the power could be increased by more than a factor of two before reaching the ionization thresold. Alternatively, for faster transfer between the ground-state hyperfine sublevels, we can use two-photon optical Raman transition \cite{Knoernschild2010,Jau2016,Levine2022}.

\section{Measurement and error correction cycle time}
\label{sec.time}

We proceed to estimate the time needed to measure and reset ancilla states in a multi-qubit array using the multi-atom mapping protocol described above. 
The number of photoelectrons generated in a measurement time $t_{\rm m}$ for an atom in the bright state $\ket{1}_{\rm m}$ is 
\begin{equation}
q(t_{\rm m})= r_s \left( \eta \frac{\Omega_{\rm d}}{4\pi}\right) t_{\rm m},
\label{eq.perate}
\end{equation}
where  
\be
r_s=\frac{\gamma }{2}\frac{I/I_s}{1+4\Delta^2/\gamma^2+I/I_{\rm s}} ~(\rm s^{-1}).
\label{eq.rs}
\ee
is the atomic scattering rate which depends on the intensity of the readout light $I$ (approximated by a sum over all beams), the atomic saturation intensity $I_{\rm s}$, the detuning $\Delta$ from the atomic transition resonance, and the decay rate $\gamma$ of the atomic excited state $\ket{f}_\mathrm{m}$. The factor in parentheses in Eq.~(\ref{eq.perate}) determines the efficiency with which scattered photons are converted into detected photoelectrons. The efficiency factor depends on the solid angle of the light collection optics $\Omega_{\rm d}$, and the quantum efficiency $\eta=\eta_{\rm d} \eta_{\rm loss}$, which accounts for  the detector efficiency $\eta_{\rm d}$, and any optical losses $\eta_{\rm loss}$.

When the atom is in the dark state $\ket{0}_\mathrm{m}$, it does not scatter any photons but photoelectron counts   accumulate   due to detector dark rate and background light scattering. Assuming negligible background light scattering, we can write for the mean background count $b(t_{\rm m})=b_0 t_{\rm m}$ with $b_0$ due to detector dark counts.  When the atoms are in the bright state $\ket{1}_\mathrm{m}$, the mean \rsub{detected} signal is $s(t_{\rm m})=k q(t_{\rm m})+b(t_{\rm m})$ for $k$ measuring atoms.  
After the exposure, photoelectrons are converted to camera counts by digitizing the voltage of the well such that the ratio of camera counts to photoelectrons is given by a calibration factor $\kappa$. 
\rsub{The critical factor for reaching high measurement fidelity is the ratio of signal to background counts, $s(t_{\rm m})/b(t_{\rm m})=k[q(t_{\rm m})/b(t_{\rm m})] + 1$, which, in the limit of small background $b(t_{\rm m}) \ll q(t_{\rm m})$, is proportional to the number of measurement atoms $k$. 
We can equivalently state that the measurement time $t_{\rm m}$ to reach a desired fidelity scales inversely with $k$.}

To perform the measurement, we illuminate the atoms and accumulate photoelectrons for a time $t_{\rm m}$ from $N_{\rm p}=9$ pixels in  a $3 \times 3$ square region on the camera. Both the signal and background photoelectrons are assumed Poisson distributed so that the probability of generating $n$ photoelectrons is
$P_n(r) = e^{-r}r^n/n!$ with $\sum_{n=0}^\infty n P_n(r)=r$ being the mean  rate.
If the number of resulting camera counts is greater than or equal to a cutoff number $n_{\rm t}$, we conclude that the atom is in state $\ket{1}$. The measurement is incorrect if the atom was in bright state $\ket{1}$ and the number of counts was less than $n_{\rm t}$ or the atom was in dark state $\ket{0}$ and the number of camera counts was greater than or equal to $n_{\rm t}$. We note that the measurement fidelity can be improved beyond what is possible with this type of thresholding measurement using more sophisticated processing of the camera counts \cite{Phuttitarn2023}.

\begin{figure}[!t]
\includegraphics[width=1.0\columnwidth]{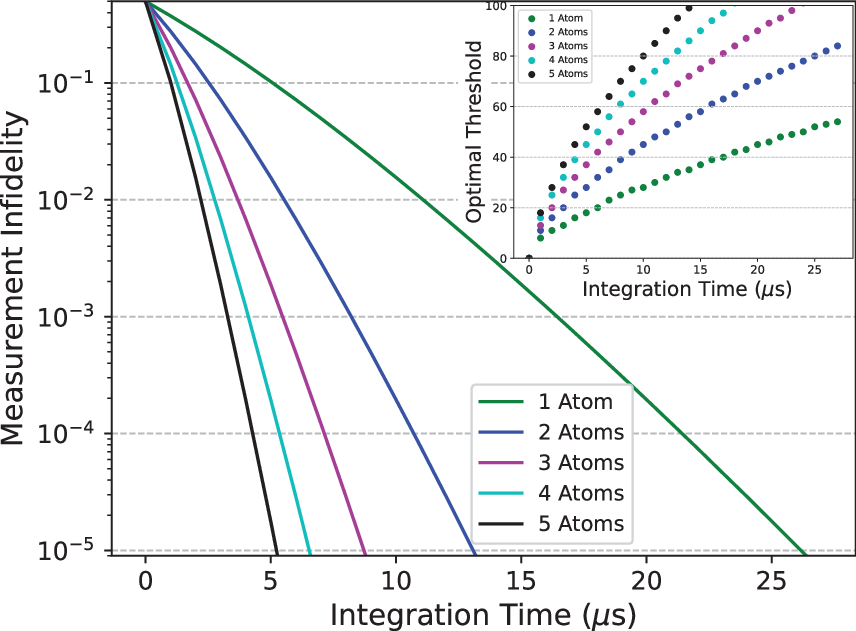}
\caption{State measurement infidelity as a function of integration time and the number of measurement atoms.
The inset shows the optimal camera count threshold values. 
 Simulation parameters were used corresponding to measurement of Rb atoms at 780 nm with an ORCA-Quest qCMOS camera: 
$\Omega_d/4\pi=0.14,$ $\Delta/\gamma=-1/2,$ $I/I_s=5,$ $\gamma=2\pi \times 6.06\times 10^6  ~\rm (s^{-1}),$ $\eta_{d}=0.54$, $\eta_{\rm loss}=0.9$, $b_0=0.016  ~\rm (s^{-1}/pixel),$ $\sigma=4 ~\rm camera~ counts/pixel$ (assuming standard readout mode), $\kappa=1/0.107,$ and $P_{\rm dark}=P_{\rm bright}=1/2.$ With these parameters $q/t_{\rm m}=9.25\times 10^5  ~\rm (s^{-1})/atom$.}
\label{fig.camera}
\end{figure}

For devices such as EMCCD cameras that convert photoelectrons to camera counts with very low noise, sums over $P_n$ can be evaluated analytically to give closed form expressions for the measurement error \cite{Saffman2005a}. We consider here the use of a qCMOS camera that  adds   normally distributed camera readout noise with mean $\mu=0$ and standard deviation $\sigma$ to the signal from each  pixel. 
The probability of a measurement error is therefore 
\begin{eqnarray}
\nonumber E(\tau) &=&  \frac{P_{\rm dark}}{2}\sum_{n=0}^\infty P_n(b)  \left[ 1-\mathrm{erf}\left(\frac{n_{\rm t}- n \kappa}{\sqrt{2 N_{\rm p}}\sigma}\right)\right] \\ 
&+ &\frac{P_{\rm bright}  }{2}\sum_{n=0}^{\infty} P_{n}(s)  \left[ 1 + \mathrm{erf}\left(\frac{n_{\rm t}-n \kappa }{\sqrt{2 N_{\rm p}}\sigma}\right)\right] 
\end{eqnarray}
where $P_{\rm dark}, P_{\rm bright}$ are the probabilities that the atom is in the dark or bright states.
For given values of $t_{\rm m}$, $q(t_{\rm m})$, $b(t_{\rm m})$, $\sigma$, and $\kappa$, there is an optimum choice of $n_{\rm t}$ that minimizes the error. 
Figure \ref{fig.camera} shows measurement error as a function of integration time for up to $k=5$ atoms. The calculated curves assume the background noise is entirely due to camera electronics and not background optical scattering. This is the case using a geometry with readout beams propagating in a plane perpendicular to the axis of the collection lens \cite{Graham2023b}. We see that with 5 measurement atoms an infidelity less than $10^{-4}$ is possible with an integration time $t_\mathrm{m} < 5~\mu\rm s$. \rsub{It should be noted that the required integration time for a desired measurement fidelity depends on the photoelectron generation rate which in turn is a function of the intensity and detuning of the readout light, as given by Eq. (\ref{eq.rs}). The readout light parameters also determine the equilibrium atomic temperature and for the parameters of Fig. \ref{fig.camera} a simplified one-dimensional theory \cite{Lett1989} predicts an equilibrium atom temperature of $510 ~\mu\rm K$. However, for measurement times of $t_{\rm m} \le 25~\mu\rm s$ the scattering induced heating amounts to $\le 40~\mu\rm K$ in each of three dimensions. Thus fast measurements on a $\mu\rm s$ timescale are consistent with low atom loss using trap depths of a few hundred $\mu\rm K$, as has been demonstrated recently \cite{LSu2024}.  }

The cycle time for a single round of syndrome measurements of surface code depends on three main contributions: a) the time needed to perform the gates that map syndrome information from data to ancilla qubits, b) the ancillae measurement time, and c) the ancillae and measurement atom reset time. A round of data - ancilla, $\sf X$ and $\sf Z$ syndrome mapping gates can be implemented as two Hadamard gates on the ancillae, two Hadamard gates on the data, and eight ancilla-data $\sf CZ$ gates \cite{Versluis2017}.   Assuming  Hadamard and $\sf CZ$ gate times of $0.5~\mu\rm s$ and simultaneous application of Hadamard gates to data and ancilla in a two-species architecture, we have a total of $t_{\rm a}=5~\mu\rm s$ for the gates.  

The measurement time, assuming five measurement atoms is $1~\mu\rm s$ for the two Rydberg pulses on the ancilla atoms at a $10 ~\rm MHz$ ground-Rydberg Rabi rate, $5~\mu\rm s$ for the state transfer of the measurement atoms shown in Fig. \ref{fig:Rb01trdyn}, and $5~\mu\rm s$ for camera integration. The time needed to readout the camera count values must also be included. This time depends on the particular camera or imaging detector used. For the ORCA-Quest qCMOS camera in standard readout mode  the minimum exposure time is $7.2~\mu\rm s$ and the 
readout time per frame assuming 128 pixel rows is $500~\mu\rm s$. Assuming negligible processing time once the camera is readout, we have a total of  
$t_b=512~\mu\rm s$ including readout time and $t_b'=12~\mu\rm s$ excluding the readout time. 

After the measurement, the ancilla and measurement atoms should be reset. The measurement result determines the ancilla state, which either is $\ket{0}_{\rm a}$ and does not need resetting, or is $\ket{1}_{\rm a}$ and can be reset with a $\pi$ pulse of duration $1~\mu\rm s$, giving an average ancilla reset time of $0.5~\mu\rm s$. The measurement atoms are either in state $\ket{0}_{\rm m}$ and do not need resetting or mostly in state $\ket{1}_{\rm m}$ if the measurement result indicated a bright state. Resetting can then be done by optical pumping into $\ket{1}_{\rm m}$ with $\sigma_+$ polarized light fields driving the transitions $\ket{5s_{1/2}, f=1}\rightarrow \ket{5p_{3/2},f=2}$ and  $\ket{5s_{1/2}, f=2}\rightarrow \ket{5p_{3/2},f=2}$ which will pump atoms into $\ket{1}_{\rm m}=\ket{5s_{1/2},f=2,m_f=2}$ which is dark to the pumping light. The measurement atoms can then be transferred to $\ket{0}_{\rm m}$ with a resonant $\pi$ pulse. Based on the rate equation simulations, we estimate that pumping and resetting of the measurement atoms can be accomplished in $10~\mu\rm s$ giving an average reset time of $5~\mu\rm s$. The total time for step c) is $t_c=5~\mu\rm s$ since the dual-species ancilla and measurement atoms can be reset simultaneously. 

Combining the time budget for each step, we find a total cycle time of $t_{\rm cycle}=522~\mu\rm s$ or $t_{\rm cycle}'=22~\mu\rm s$ excluding the camera data transfer time. With the assumption of the availability of proposed optical devices for fast generation of multi-spot light patterns \cite{Graham2023,Menssen2023,BZhang2024}, these timing estimates apply to arrays   of thousands of physical qubits. In order to advance from the $\sim 2~\rm kHz$ rate  limited by the data transfer time, to rates approaching $50~\rm kHz$, it will be necessary to incorporate high speed interfaces to multi-pixel, low-noise detectors \cite{Ulku2019}. 

\section{Multiqubit ${\sf CNOT}_k$ gate}
\label{sec.gate}

Neutral atom qubits provide a natural setting for efficient implementation of multi-qubit gates using the strong and long range nature of the Rydberg blockade mechanism which enabled several early proposals for multi-qubit interactions \cite{Brion2007c, Isenhower2011, Petrosyan2016}. A large number of proposals for multi-qubit Rydberg gates  have emerged recently 
\cite{Su2018,
XFShi2018,
Beterov2018b, 
Khazali2020,
Young2021,
Rasmussen2020,
THXing2020,  
JLWu2021, 
YHe2022,
MLi2021,
Pelegri2022,
Farouk2022,
Kinos2023}, as well as an experimental  demonstration of a  3-qubit Toffoli gate \cite{Levine2019}, and a 9-qubit generalized phase gate \cite{ACao2024}. 

\begin{figure}[!t]
\includegraphics[width=1.0\columnwidth]{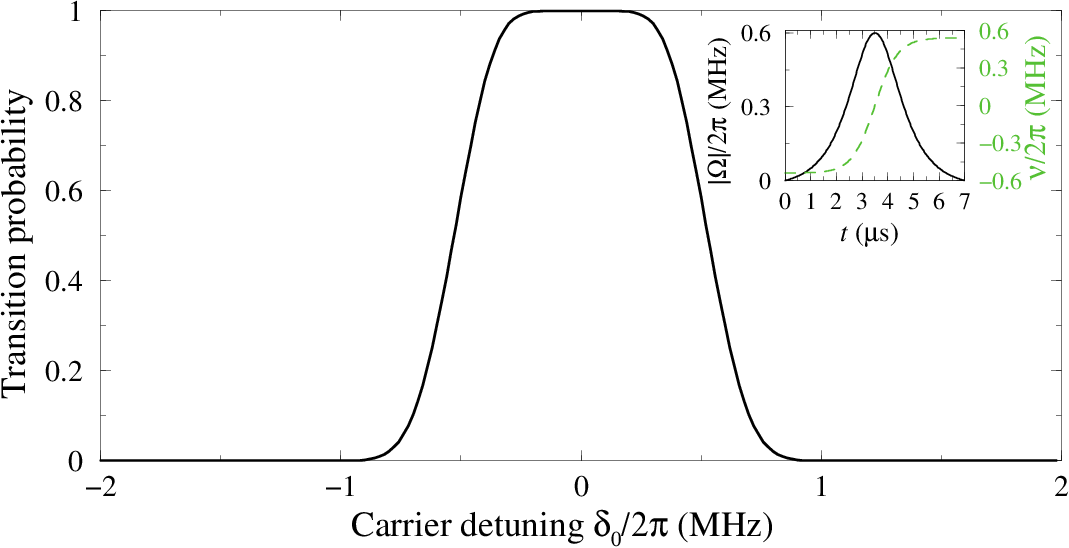}
\caption{Transition probability $\braket{\sigma_{11}(t=\tau)}$ between states $\ket{0}$ and $\ket{1}$ for a two-level atom driven by a chirped adiabatic pulse $\Omega(t)= \Omega_\mathrm{max} \{ \mathrm{sech}[\beta (t-\tau/2)] \}^{1+i \mu}$ with $\Omega_\mathrm{max} = 2\pi \times 0.6\,$MHz, 
$\beta = 2\pi \times 0.18\,$MHz and $\mu=3$, vs the carrier detuning $\delta_0$.
Inset illustrates the pulse amplitude $|\Omega(t)|= \Omega_\mathrm{max} \mathrm{sech}[\beta (t-\tau/2)]$ and frequency chirp  $\nu(t) = \mu \beta \, \mathrm{tanh}[\beta (t-\tau/2)]$ corresponding to the detuning sweep $\delta(t) = \delta_0 + \nu(t)$ around $\delta_0$.}
\label{fig:tlaAp}
\end{figure}

We note that the present protocol amounts to performing the $\textsf{CNOT}_k$ gate between one Cs control qubit and $k\geq 1$ Rb target qubits, if the transfer between the states $\ket{0}_\mathrm{m}$ and $\ket{1}_\mathrm{m}$ of the latter, or the suppression thereof, can be made with high probability and well-defined phase. 
Ideally, robust transfer, or complete inhibition thereof, of the target qubits between their  basis states should be immune to small perturbations of the atomic positions and field amplitudes. This means that the transition spectrum should have a bowler-hat shape with a flat top around the resonance, steep edges and zero background away from resonance, see Fig.~\ref{fig:tlaAp}. 
This can be achieved using, e.g., a chirped adiabatic pulse 
\begin{equation}
\Omega(t)= \Omega_\mathrm{max} \{ \mathrm{sech}[\beta (t-\tau/2)] \}^{1+i \mu} \label{eq:OmegaChirped}
\end{equation}
having the amplitude 
$|\Omega(t)|= \Omega_\mathrm{max} \mathrm{sech}[\beta (t-\tau/2)]$ and frequency chirp  
$\nu(t) = \mu \beta \, \mathrm{tanh}[\beta (t-\tau/2)]$
around the carrier detuning $\delta_0$ \cite{Roos2004}, as illustrated in the Inset of Fig.~\ref{fig:tlaAp}. 
The parameters of this pulse have to satisfy the following requirements:
the bandwidth $\mu \beta$ should be, on the one hand, large enough to accommodate the level shifts for a range of interaction strengths $B_{0j}$ between the control (Cs) and target (Rb) atoms in Rydberg states $\ket{r}_{\mathrm{a,m}}$ at relative position uncertainty $\Delta r_{\mathrm{am}} \simeq 40\:$nm ($\simeq 1\%$ of $ r_{\mathrm{am}}$ and $\simeq 5\%$ variation of $B_{0j}$); and, on the other hand, smaller than the ac Stark shift for non-interacting atoms (Cs in state $\ket{0}$), $\frac{|\Omega_r|^2}{4 (\Delta_r+B_{0j})} < \mu \beta < \frac{|\Omega_r|^2}{4 \Delta_r}$. 
Simultaneously, the pulse duration $\tau \sim 10/\beta$ should
be as short as possible to minimize the decay probability $\gamma_{r_{\rm a}} \tau$ of the Rydberg state of the Cs atoms during the transfer.
Finally, the pulse amplitude should be sufficiently large for the transfer to be adiabatic, i.e., vanishingly small non-adiabatic (Landau-Zener) transition probability between the adiabatic eigenstates,  
$\exp\left[-2\pi \frac{|\Omega_\mathrm{max}/2|^2}{\partial_t \nu(t)} \right] \to 0$, which leads to the condition $2\pi \frac{|\Omega_\mathrm{max}|^2}{4 \mu \beta^2} \gg 1$ and therefore $\Omega_{\mathrm{max}} > 2\beta \sqrt{\mu/2\pi}$.

To satisfy these conditions, we assume the Rydberg dressing laser with $\Omega_r/2\pi = 10\:$MHz and $\Delta_r/2\pi = 25\:$MHz with a  Cs-Rb interaction strength $B_{0j} \gtrsim 4 \Delta_r$, leading to $0.2 < \mu \beta/2\pi < 1\:$MHz. With $\mu =3$ and $\beta/2\pi = 0.18\:$MHz, we set the pulse duration $\tau = 7\:\mu$s and its peak Rabi frequency $\Omega_\mathrm{max}/2\pi = 0.6\:$MHz so that the exponent in the Landau-Zener formula is $\simeq 5.8$ and the nonadiabatic transition probability is less than $3\times 10^{-3}$.     
Such a large Rabi frequency for the magnetic dipole transition $\ket{0}_\mathrm{m} \leftrightarrow \ket{1}_\mathrm{m}$ of the Rb atoms would require the microwave field amplitude 
$E_{\rm max} \gtrsim 10^4\:$V/m which is much stronger than the ionization field for the Rydberg states $\ket{r}_\mathrm{a} \equiv \ket{64s_{1/2},m_j=1/2}$ of Cs and $\ket{r}_\mathrm{m} \equiv \ket{60s_{1/2},m_j=1/2}$ of Rb that we now consider.
Instead, we assume a two-photon optical Raman transition between the ground-state hyperfine sublevels $\ket{0}_{\rm m} \equiv \ket{5s_{1/2}, f=1,m_f=0}$ and $\ket{1}_{\rm m} \equiv \ket{5s_{1/2},f=2,m_f=0}$ via the intermediate non-resonant state $\ket{5p_{1/2}, f=1,m_f=1}$. 
\rsub{Note that for atoms in traps with the motional degrees of freedom cooled to $1-10\:\mu$K temperatures, the Doppler shift on the optical transition $\Delta_{\mathrm{Doppler}} \simeq 2\pi \times 10-40~$kHz is much smaller than the detuning $\Delta_r$ of the Rydberg dressing laser and can therefore be safely neglected.
Simultaneously, the Doppler shift on the Raman transition between the ground-state sublevels is vanishingly small and does not affect the bowler-hat shaped transition spectrum of the Rb atoms when the Raman lasers are copropagating.}

\begin{figure}[!t]
\includegraphics[width=1.0\columnwidth]{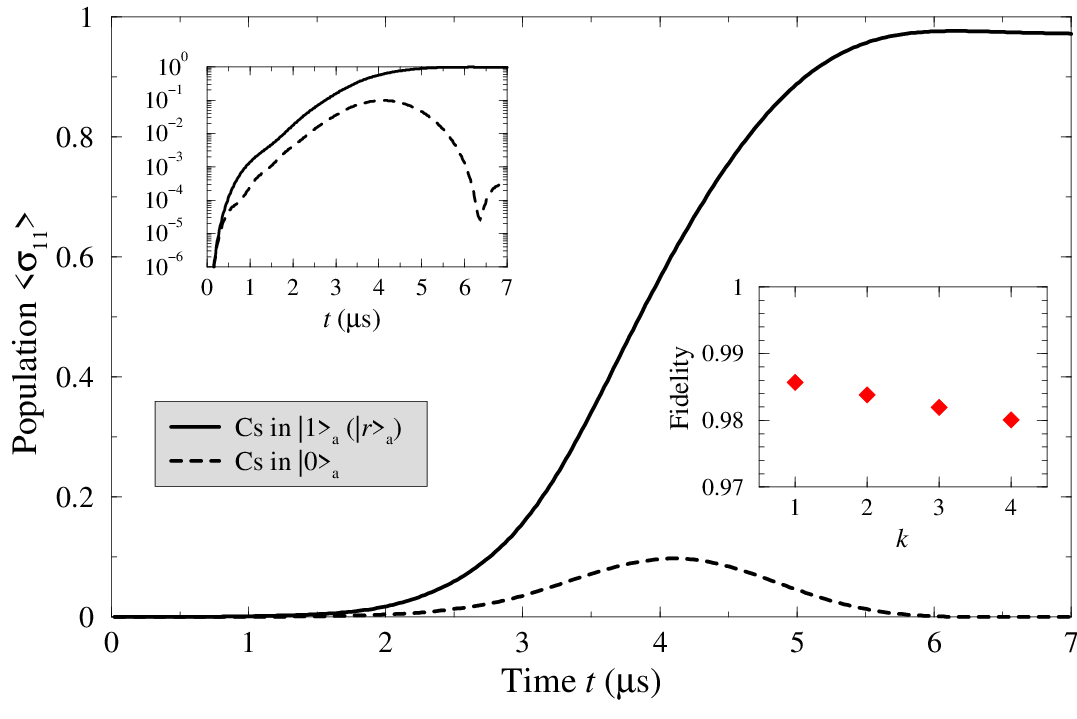}
\caption{Dynamics of populations $\braket{\sigma_{11}^{(j)}}$ of state $\ket{1}_{\rm m}$ of each of the target Rb atoms, for $k=1,2,3,4$ atoms (nearly indistinguishable overlapping curves), initially in state $\ket{0}_{\rm m}$, upon applying a chirped adiabatic pulse $\Omega(t)$ of Fig.~\ref{fig:tlaAp}.
We assume the $\ket{r}_\mathrm{a} \equiv \ket{64s_{1/2},m_j=1/2}$ Rydberg state of Cs with  lifetime $1/\gamma_{r_{\rm a}}= 280\:\mu$s, and the $\ket{r}_\mathrm{m} \equiv \ket{60s_{1/2},m_j=1/2}$ Rydberg state of Rb with  lifetime $1/\gamma_{r_{\rm m}}= 250\:\mu$s at temperature $T=4\:$ K. 
The interaction strength between the Cs and Rb atoms separated by $r_{\mathrm{am}} =4\:\mu$m is $B_{0j} = 2\pi \times 110\:$ MHz, while the interaction strengths between the Rb atoms on a circle are $B_{12} = 2\pi \times 0.8\:$ MHz for $k=2$, $B_{12,23,13} = 2\pi \times 1.6\:$ MHz for $k=3$, and  $B_{12,23,34,14} = 2\pi \times 4.7\:$ MHz for $k=4$. 
For the control Cs atom in Rydberg state $\ket{r}_\mathrm{a}$, the target Rb atoms undergo population transfer to state $\ket{1}_{\rm m}$ (solid lines); while for the control Cs atom in state $\ket{0}_\mathrm{a}$, the transfer is suppressed (dashed lines). 
Left upper inset shows the same populations on a log scale.
Right lower inset shows the preparation fidelity 
$\mathcal{F}=|\braket{\mathrm{GHZ}|\Psi(\tau)}|^2$
of the GHZ state of $k+1$ atoms with the ${\sf CNOT}_k$ gate involving one control (Cs) and $k$ target (Rb) atoms.
About 1.2\% of the fidelity is lost due to the decay of the Rydberg state of the Cs atom, the remaining decrease is due to the decay of the Rydberg state of each of the Rb atoms which is suppressed by a factor of $|\Omega_r|^2/4\Delta_r^2$. }
\label{fig:Rbgate01trdyn}
\end{figure}

With these parameters, the nearly complete transfer, $\braket{\sigma_{11}^{(j)}} \to 1$ for $\ket{1}_\mathrm{a} \to \ket{r}_\mathrm{a}$, or suppression thereof, $\braket{\sigma_{11}^{(j)}} \to 0$ for $\ket{0}_\mathrm{a}$, controlled by the state $\ket{\psi}_\mathrm{a} \in \{ \ket{0}_\mathrm{a} , \ket{1}_\mathrm{a} \}$ of the Cs atom,
is then possible, as we show in Fig.~\ref{fig:Rbgate01trdyn}, as obtained from  numerical simulations. 
The main error source that we encounter is a rather large decay probability $\gamma_{r_{\rm a}} \tau \simeq 0.025$ of the Rydberg state of the Cs atom during the transfer time $\tau$, even when assuming cryogenic temperatures $T=4\:$K. This problem can be mitigated by going to even higher Rydberg states $n=80-100$ with longer lifetimes $1/\gamma_r \propto n^3 \simeq 0.5-1\:$ms.

To quantify the performance of the ${\sf CNOT}_k$ gate, we use the present protocol to prepare a GHZ state of $k+1$ qubits and calculate the fidelity of the state preparation. To that end, we first prepare the control (Cs) qubit in the coherent superposition state $\ket{\psi}_\mathrm{a} = \frac{1}{\sqrt{2}} (\ket{0}_\mathrm{a} + \ket{1}_\mathrm{a})$, while all $k$ target (Rb) qubits are initialized in state $\ket{0}_\mathrm{m}$. The initial state of the system is then $\ket{\psi}_\mathrm{a} \otimes \ket{0}_\mathrm{m}^{\otimes k}$. Next, we transfer the control qubit state $\ket{1}_\mathrm{a} \to \ket{r}_\mathrm{a}$, apply the (Raman) pulse of Eq.~(\ref{eq:OmegaChirped}) to all the target qubits, and return the control qubit state $\ket{r}_\mathrm{a} \to \ket{1}_\mathrm{a}$. The resulting evolution of the compound state $\ket{\Psi}$  is then
\begin{eqnarray}
\ket{\Psi(0)} &=& \frac{\ket{0}_\mathrm{a} + \ket{1}_\mathrm{a}}{\sqrt2} \otimes \ket{0}_\mathrm{m}^{\otimes k}\nonumber\\
&\to& \frac{ \ket{0}_\mathrm{a} \otimes \ket{0}_\mathrm{m}^{\otimes k} + e^{\imath k\phi_0} \ket{1}_\mathrm{a} \otimes \ket{1}_\mathrm{m}^{\otimes k}}{\sqrt2} \equiv\ket{\mathrm{GHZ}} ~~~
\end{eqnarray}
where $\phi_0 = 2.03\:$rad is the single-atom dynamical phase accumulated during the transfer.
We simulate the dynamics of the complete system, composed of one control qubit and $k=1,2,3,4$ target qubits,  
governed by the Hamiltonian (\ref{eq:ham}). 
The resulting fidelity 
$\mathcal{F} = |\braket{\mathrm{GHZ}|\Psi(\tau)}|^2$
of the GHZ (or Bell, for $k=1$) state is shown in the inset of Fig.~\ref{fig:Rbgate01trdyn}.
We obtain $\mathcal{F} \geq 0.98$ for all $k \leq 4$.
The main factor that reduces the fidelity is the decay of the Rydberg state $\ket{r}_{\mathrm{a}}$ of the Cs atom during the long gate time $\tau$. Then the fidelity decreases with increasing the number of target atoms due to the decay of the Rydberg state $\ket{r}_\mathrm{m}$ with which we dress the qubit state $\ket{1}_\mathrm{m}$ and therefore lose during the transfer some population from that state, $\Delta \braket{\sigma_{11}^{(j)}} \lesssim \frac{1}{2} \gamma_{r} \tau \frac{|\Omega_r|^2}{4\Delta_r^2}$.
Again, going to higher Rydberg states for both Cs and Rb, and/or using the longer-lived $nP$ states, will increase the gate fidelity, and multiqubit Rydberg gates with error probability of $<1\%$ appear feasible under realistic experimental conditions. 
\rsub{We  note finally that in our simulations we neglected dephasing of the Rydberg state of Cs atoms and qubit states of the Cs and Rb atoms due to, e.g., external electric or magnetic field noise or laser phase fluctuations, since these are technical imperfections that should be suppressed in any experiment intended to achieve high fidelity quantum gates with atomic qubits.}

\section{Conclusions}
\label{sec.outlook}

To conclude, we have suggested and analysed a scheme to perform syndrome measurements for quantum computation and error correction in a dual-species atomic qubit array. The proposal offers decreased measurement time over existing proposals, with robust crosstalk-free measurements of ancilla qubits. With the state of each ancilla atom mapped to five auxiliary measurement atoms, quantum error correction cycle times close to $20~\mu\rm s$ appear feasible. This estimate includes the time needed for syndrome measurement gates, the atomic state measurement, and reset operations on the ancilla and measurement atoms, but neglects the time required for data transfer from the camera used for measurements. Currently available cameras are compatible with quantum error correction cycle rates of 2 kHz, and close to 50 kHz appears feasible with faster camera readout. 
We finally note that the proposed scheme natively provides an inter-species ${\sf CNOT}_k$ gate between one Cs control qubit and $k\geq1$ Rb target qubits. 

\rsub{
Note added: After submission of this work related results appeared in \cite{Corlett2024}.}

\acknowledgments

This work was supported by
NSF Award 2210437,  
NSF Award 2016136 for the QLCI Hybrid Quantum Architectures and Networks,
the
U.S. Department of Energy Office of Science National
Quantum Information Science Research Centers as part of the Q-NEXT center, 
NSF Award 2228725, and Infleqtion, Inc. 
D.P. was supported by the EU HORIZON-RIA Project EuRyQa (grant No. 101070144).

\appendix*

\section{Rydberg state parameters and calculation methodology}

The lifetimes  $\tau_{r}=1/\gamma_r$ and decay rates $\gamma_{r}$ of the Rydberg states used in the main text are given in Table \ref{tab.lifetime} for several temperatures \cite{Beterov2009}. Direct calculation of decay rates using a sum over states with radial matrix elements calculated from the Alkali Rydberg Calculator (ARC) program \cite{ARC3.0} resulted in lifetimes approximately 10\% shorter than those  in Table~\ref{tab.lifetime}.

\begin{table}[h]
\caption{
Rydberg state lifetimes and decay rates. 
}
\label{tab.lifetime}
\centering
\begin{tabular}{|c| c| c|c|c|c|c|c|c|c|c|}
\hline
 $T~(\rm K)$ & state & $\tau_{r}~(\mu\rm s)$ & $\gamma_{r}~(10^3 \rm s^{-1})$ & State & $\tau_{r}~(\mu\rm s)$ & $\gamma_{r}~(10^3 \rm s^{-1})$    \\
\hline
300. & Rb 46s  & 52.1 & 19.2  & Cs 48s  & 53.8 & 18.6 \\
77 & Rb 46s  & 85.2 & 11.7   & Cs 48s  & 87.2 & 11.5\\
4 & Rb 46s  & 107 & 9.32   & Cs 48s & 109 & 9.16\\
\hline
300. & Rb 60s & 101 & 9.90  & Cs 64s  & 112 & 8.93  \\
77 & Rb 60s  & 183 & 5.46   & Cs 64s  & 202 & 4.95\\
4 & Rb 60s  & 249 & 4.02   & Cs 64s  & 275 & 3.64  \\
\hline
\end{tabular}
\end{table}

\begin{figure}[t!]
\includegraphics[width=0.9\columnwidth]{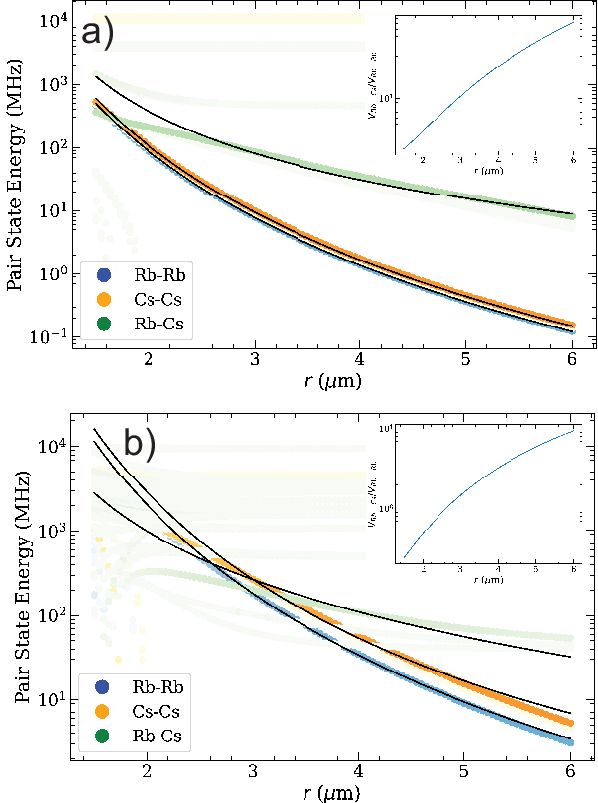}
\caption{Interaction potentials for the Rydberg pair states (a) Rb46s - Cs48s, and (b) Rb60s - Cs64s. Colored points are the relative energies of eigenstates of the diagonalized Hamiltonian (shading is proportional to the eigenstate overlap $g$ with the target state). Solid lines are the fits used to extract the $C_3$ and $C_6$ coefficients in Table~\ref{table.c3c6-extracted}. Insets show the ratio of inter-species Rb-Cs interaction energy to the intra-species Rb-Rb interaction energy.
}
\label{fig.C3C6}
\end{figure}

Atomic pair state interaction strengths were calculated with ARC \cite{ARC,ARC3.0}. Pair states were calculated in a $\theta = \frac{\pi}{2}, \phi=0$ geometry, with $m_j=1/2$. The interaction Hamiltonian was constructed from a range of states specified by $\Delta n = \pm 2, \Delta l = \pm 2$. 
For each Hamiltonian (Rb-Rb, Cs-Cs, and Rb-Cs), 250 eigenstates were extracted for the inter-atomic separation distances $r=1.5 -6\;\mu$m. 
For each case, the arithmetic mean of the pair interaction curves, weighted by the overlap of the eigenstate with the target state $g=|\braket{\psi_{\rm pair-state}|\psi_{\rm target}}|^2$, was fitted by a function of the form $V(r)=\frac{C_3}{r^3}+ \frac{C_6}{r^6}$. Only states with $g\geq 0.4$ for Rb-Rb and Cs-Cs and $g\geq 0.01$ for Rb-Cs were included for fitting. The fits were performed over the interval $3~ \mu$m $\leq r \leq 6~ \mu$m, and results are presented in Fig.~\ref{fig.C3C6} and Table~\ref{table.c3c6-extracted}. \rsub{For visual clarity, discrete points in Fig.~\ref{fig.C3C6} with $g<0.05$ are not shown.}

\begin{table}[h]
\caption{
Pair state interaction coefficients $C_3$ and $C_6$ in units of ${\rm GHz}~ \mu\rm m^3$ and ${\rm GHz}~ \mu\rm m^6$.  }
\label{table.c3c6-extracted}
\begin{tabular}{|l|c|c|c|c|c|c|c|c|}
\hline
Pair State  & Rb-Rb  & Rb-Rb  & Cs-Cs & Cs-Cs & Rb-Cs & Rb-Cs   \\
  & $C_3$   &  $C_6$ &  $C_3$&  $C_6$ &  $C_3$ &  $C_6$  \\
\hline
Rb46s-Cs48s & 0.000108  & 5.65  & 0.000341    & 6.86   & 1.87 & 8.867     \\
Rb60s-Cs64s & 0.1388 & 129.8 & 0.664     & 178.7 & 6.955  & 7.466  \\
\hline
\end{tabular}
\end{table}


\bibliography{atomic,saffman_refs,rydberg,qc_refs,optics,norrell-refs}

\begin{thebibliography}{55}%
\makeatletter
\providecommand \@ifxundefined [1]{%
 \@ifx{#1\undefined}
}%
\providecommand \@ifnum [1]{%
 \ifnum #1\expandafter \@firstoftwo
 \else \expandafter \@secondoftwo
 \fi
}%
\providecommand \@ifx [1]{%
 \ifx #1\expandafter \@firstoftwo
 \else \expandafter \@secondoftwo
 \fi
}%
\providecommand \natexlab [1]{#1}%
\providecommand \enquote  [1]{``#1''}%
\providecommand \bibnamefont  [1]{#1}%
\providecommand \bibfnamefont [1]{#1}%
\providecommand \citenamefont [1]{#1}%
\providecommand \href@noop [0]{\@secondoftwo}%
\providecommand \href [0]{\begingroup \@sanitize@url \@href}%
\providecommand \@href[1]{\@@startlink{#1}\@@href}%
\providecommand \@@href[1]{\endgroup#1\@@endlink}%
\providecommand \@sanitize@url [0]{\catcode `\\12\catcode `\$12\catcode `\&12\catcode `\#12\catcode `\^12\catcode `\_12\catcode `\%12\relax}%
\providecommand \@@startlink[1]{}%
\providecommand \@@endlink[0]{}%
\providecommand \url  [0]{\begingroup\@sanitize@url \@url }%
\providecommand \@url [1]{\endgroup\@href {#1}{\urlprefix }}%
\providecommand \urlprefix  [0]{URL }%
\providecommand \Eprint [0]{\href }%
\providecommand \doibase [0]{https://doi.org/}%
\providecommand \selectlanguage [0]{\@gobble}%
\providecommand \bibinfo  [0]{\@secondoftwo}%
\providecommand \bibfield  [0]{\@secondoftwo}%
\providecommand \translation [1]{[#1]}%
\providecommand \BibitemOpen [0]{}%
\providecommand \bibitemStop [0]{}%
\providecommand \bibitemNoStop [0]{.\EOS\space}%
\providecommand \EOS [0]{\spacefactor3000\relax}%
\providecommand \BibitemShut  [1]{\csname bibitem#1\endcsname}%
\let\auto@bib@innerbib\@empty
\bibitem [{\citenamefont {Deist}\ \emph {et~al.}(2022)\citenamefont {Deist}, \citenamefont {Lu}, \citenamefont {Ho}, \citenamefont {Pasha}, \citenamefont {Zeiher}, \citenamefont {Yan},\ and\ \citenamefont {Stamper-Kurn}}]{Deist2022}%
  \BibitemOpen
  \bibfield  {author} {\bibinfo {author} {\bibfnamefont {E.}~\bibnamefont {Deist}}, \bibinfo {author} {\bibfnamefont {Y.-H.}\ \bibnamefont {Lu}}, \bibinfo {author} {\bibfnamefont {J.}~\bibnamefont {Ho}}, \bibinfo {author} {\bibfnamefont {M.~K.}\ \bibnamefont {Pasha}}, \bibinfo {author} {\bibfnamefont {J.}~\bibnamefont {Zeiher}}, \bibinfo {author} {\bibfnamefont {Z.}~\bibnamefont {Yan}},\ and\ \bibinfo {author} {\bibfnamefont {D.~M.}\ \bibnamefont {Stamper-Kurn}},\ }\bibfield  {title} {\bibinfo {title} {Mid-circuit cavity measurement in a neutral atom array},\ }\href {https://doi.org/10.1103/PhysRevLett.129.203602} {\bibfield  {journal} {\bibinfo  {journal} {Phys. Rev. Lett.}\ }\textbf {\bibinfo {volume} {129}},\ \bibinfo {pages} {203602} (\bibinfo {year} {2022})}\BibitemShut {NoStop}%
\bibitem [{\citenamefont {Bluvstein}\ \emph {et~al.}(2024)\citenamefont {Bluvstein}, \citenamefont {Evered}, \citenamefont {Geim}, \citenamefont {Li}, \citenamefont {Zhou}, \citenamefont {Manovitz}, \citenamefont {Ebadi}, \citenamefont {Cain}, \citenamefont {Kalinowski}, \citenamefont {Hangleiter}, \citenamefont {Ataides}, \citenamefont {Maskara}, \citenamefont {Cong}, \citenamefont {Gao}, \citenamefont {Rodriguez}, \citenamefont {Karolyshyn}, \citenamefont {Semeghini}, \citenamefont {Greiner}, \citenamefont {Vuleti\'c},\ and\ \citenamefont {Lukin}}]{Bluvstein2024}%
  \BibitemOpen
  \bibfield  {author} {\bibinfo {author} {\bibfnamefont {D.}~\bibnamefont {Bluvstein}}, \bibinfo {author} {\bibfnamefont {S.~J.}\ \bibnamefont {Evered}}, \bibinfo {author} {\bibfnamefont {A.~A.}\ \bibnamefont {Geim}}, \bibinfo {author} {\bibfnamefont {S.~H.}\ \bibnamefont {Li}}, \bibinfo {author} {\bibfnamefont {H.}~\bibnamefont {Zhou}}, \bibinfo {author} {\bibfnamefont {T.}~\bibnamefont {Manovitz}}, \bibinfo {author} {\bibfnamefont {S.}~\bibnamefont {Ebadi}}, \bibinfo {author} {\bibfnamefont {M.}~\bibnamefont {Cain}}, \bibinfo {author} {\bibfnamefont {M.}~\bibnamefont {Kalinowski}}, \bibinfo {author} {\bibfnamefont {D.}~\bibnamefont {Hangleiter}}, \bibinfo {author} {\bibfnamefont {J.~P.}\ \bibnamefont {Ataides}}, \bibinfo {author} {\bibfnamefont {N.}~\bibnamefont {Maskara}}, \bibinfo {author} {\bibfnamefont {I.}~\bibnamefont {Cong}}, \bibinfo {author} {\bibfnamefont {X.}~\bibnamefont {Gao}}, \bibinfo {author} {\bibfnamefont {P.~S.}\ \bibnamefont {Rodriguez}}, \bibinfo {author} {\bibfnamefont
  {T.}~\bibnamefont {Karolyshyn}}, \bibinfo {author} {\bibfnamefont {G.}~\bibnamefont {Semeghini}}, \bibinfo {author} {\bibfnamefont {M.}~\bibnamefont {Greiner}}, \bibinfo {author} {\bibfnamefont {V.}~\bibnamefont {Vuleti\'c}},\ and\ \bibinfo {author} {\bibfnamefont {M.~D.}\ \bibnamefont {Lukin}},\ }\bibfield  {title} {\bibinfo {title} {Logical quantum processor based on reconfigurable atom arrays},\ }\href@noop {} {\bibfield  {journal} {\bibinfo  {journal} {Nature}\ }\textbf {\bibinfo {volume} {626}},\ \bibinfo {pages} {58} (\bibinfo {year} {2024})}\BibitemShut {NoStop}%
\bibitem [{\citenamefont {Graham}\ \emph {et~al.}(2023{\natexlab{a}})\citenamefont {Graham}, \citenamefont {Phuttitarn}, \citenamefont {Chinnarasu}, \citenamefont {Song}, \citenamefont {Poole}, \citenamefont {Jooya}, \citenamefont {Scott}, \citenamefont {Scott}, \citenamefont {Eichler},\ and\ \citenamefont {Saffman}}]{Graham2023b}%
  \BibitemOpen
  \bibfield  {author} {\bibinfo {author} {\bibfnamefont {T.~M.}\ \bibnamefont {Graham}}, \bibinfo {author} {\bibfnamefont {L.}~\bibnamefont {Phuttitarn}}, \bibinfo {author} {\bibfnamefont {R.}~\bibnamefont {Chinnarasu}}, \bibinfo {author} {\bibfnamefont {Y.}~\bibnamefont {Song}}, \bibinfo {author} {\bibfnamefont {C.}~\bibnamefont {Poole}}, \bibinfo {author} {\bibfnamefont {K.}~\bibnamefont {Jooya}}, \bibinfo {author} {\bibfnamefont {J.}~\bibnamefont {Scott}}, \bibinfo {author} {\bibfnamefont {A.}~\bibnamefont {Scott}}, \bibinfo {author} {\bibfnamefont {P.}~\bibnamefont {Eichler}},\ and\ \bibinfo {author} {\bibfnamefont {M.}~\bibnamefont {Saffman}},\ }\bibfield  {title} {\bibinfo {title} {Mid-circuit measurements on a neutral atom quantum processor},\ }\href@noop {} {\bibfield  {journal} {\bibinfo  {journal} {Phys. Rev. X}\ }\textbf {\bibinfo {volume} {13}},\ \bibinfo {pages} {041051} (\bibinfo {year} {2023}{\natexlab{a}})}\BibitemShut {NoStop}%
\bibitem [{\citenamefont {Ma}\ \emph {et~al.}(2023)\citenamefont {Ma}, \citenamefont {Liu}, \citenamefont {Peng}, \citenamefont {Zhang}, \citenamefont {Jandura}, \citenamefont {Claes}, \citenamefont {Burgers}, \citenamefont {Pupillo}, \citenamefont {Puri},\ and\ \citenamefont {Thompson}}]{SMa2023}%
  \BibitemOpen
  \bibfield  {author} {\bibinfo {author} {\bibfnamefont {S.}~\bibnamefont {Ma}}, \bibinfo {author} {\bibfnamefont {G.}~\bibnamefont {Liu}}, \bibinfo {author} {\bibfnamefont {P.}~\bibnamefont {Peng}}, \bibinfo {author} {\bibfnamefont {B.}~\bibnamefont {Zhang}}, \bibinfo {author} {\bibfnamefont {S.}~\bibnamefont {Jandura}}, \bibinfo {author} {\bibfnamefont {J.}~\bibnamefont {Claes}}, \bibinfo {author} {\bibfnamefont {A.~P.}\ \bibnamefont {Burgers}}, \bibinfo {author} {\bibfnamefont {G.}~\bibnamefont {Pupillo}}, \bibinfo {author} {\bibfnamefont {S.}~\bibnamefont {Puri}},\ and\ \bibinfo {author} {\bibfnamefont {J.~D.}\ \bibnamefont {Thompson}},\ }\bibfield  {title} {\bibinfo {title} {High-fidelity gates and mid-circuit erasure conversion in an atomic qubit},\ }\href@noop {} {\bibfield  {journal} {\bibinfo  {journal} {Nature}\ }\textbf {\bibinfo {volume} {622}},\ \bibinfo {pages} {279} (\bibinfo {year} {2023})}\BibitemShut {NoStop}%
\bibitem [{\citenamefont {Norcia}\ \emph {et~al.}(2023)\citenamefont {Norcia}, \citenamefont {Cairncross}, \citenamefont {Barnes}, \citenamefont {Battaglino}, \citenamefont {Brown}, \citenamefont {Brown}, \citenamefont {Cassella}, \citenamefont {Chen}, \citenamefont {Coxe}, \citenamefont {Crow}, \citenamefont {Epstein}, \citenamefont {Griger}, \citenamefont {Jones}, \citenamefont {Kim}, \citenamefont {Kindem}, \citenamefont {King}, \citenamefont {Kondov}, \citenamefont {Kotru}, \citenamefont {Lauigan}, \citenamefont {Li}, \citenamefont {Lu}, \citenamefont {Megidish}, \citenamefont {Marjanovic}, \citenamefont {McDonald}, \citenamefont {Mittiga}, \citenamefont {Muniz}, \citenamefont {Narayanaswami}, \citenamefont {Nishiguchi}, \citenamefont {Notermans}, \citenamefont {Paule}, \citenamefont {Pawlak}, \citenamefont {Peng}, \citenamefont {Ryou}, \citenamefont {Smull}, \citenamefont {Stack}, \citenamefont {Stone}, \citenamefont {Sucich}, \citenamefont {Urbanek}, \citenamefont {van~de Veerdonk}, \citenamefont
  {Vendeiro}, \citenamefont {Wilkason}, \citenamefont {Wu}, \citenamefont {Xie}, \citenamefont {Zhang},\ and\ \citenamefont {Bloom}}]{Norcia2023}%
  \BibitemOpen
  \bibfield  {author} {\bibinfo {author} {\bibfnamefont {M.~A.}\ \bibnamefont {Norcia}}, \bibinfo {author} {\bibfnamefont {W.~B.}\ \bibnamefont {Cairncross}}, \bibinfo {author} {\bibfnamefont {K.}~\bibnamefont {Barnes}}, \bibinfo {author} {\bibfnamefont {P.}~\bibnamefont {Battaglino}}, \bibinfo {author} {\bibfnamefont {A.}~\bibnamefont {Brown}}, \bibinfo {author} {\bibfnamefont {M.~O.}\ \bibnamefont {Brown}}, \bibinfo {author} {\bibfnamefont {K.}~\bibnamefont {Cassella}}, \bibinfo {author} {\bibfnamefont {C.-A.}\ \bibnamefont {Chen}}, \bibinfo {author} {\bibfnamefont {R.}~\bibnamefont {Coxe}}, \bibinfo {author} {\bibfnamefont {D.}~\bibnamefont {Crow}}, \bibinfo {author} {\bibfnamefont {J.}~\bibnamefont {Epstein}}, \bibinfo {author} {\bibfnamefont {C.}~\bibnamefont {Griger}}, \bibinfo {author} {\bibfnamefont {A.~M.~W.}\ \bibnamefont {Jones}}, \bibinfo {author} {\bibfnamefont {H.}~\bibnamefont {Kim}}, \bibinfo {author} {\bibfnamefont {J.~M.}\ \bibnamefont {Kindem}}, \bibinfo {author} {\bibfnamefont
  {J.}~\bibnamefont {King}}, \bibinfo {author} {\bibfnamefont {S.~S.}\ \bibnamefont {Kondov}}, \bibinfo {author} {\bibfnamefont {K.}~\bibnamefont {Kotru}}, \bibinfo {author} {\bibfnamefont {J.}~\bibnamefont {Lauigan}}, \bibinfo {author} {\bibfnamefont {M.}~\bibnamefont {Li}}, \bibinfo {author} {\bibfnamefont {M.}~\bibnamefont {Lu}}, \bibinfo {author} {\bibfnamefont {E.}~\bibnamefont {Megidish}}, \bibinfo {author} {\bibfnamefont {J.}~\bibnamefont {Marjanovic}}, \bibinfo {author} {\bibfnamefont {M.}~\bibnamefont {McDonald}}, \bibinfo {author} {\bibfnamefont {T.}~\bibnamefont {Mittiga}}, \bibinfo {author} {\bibfnamefont {J.~A.}\ \bibnamefont {Muniz}}, \bibinfo {author} {\bibfnamefont {S.}~\bibnamefont {Narayanaswami}}, \bibinfo {author} {\bibfnamefont {C.}~\bibnamefont {Nishiguchi}}, \bibinfo {author} {\bibfnamefont {R.}~\bibnamefont {Notermans}}, \bibinfo {author} {\bibfnamefont {T.}~\bibnamefont {Paule}}, \bibinfo {author} {\bibfnamefont {K.}~\bibnamefont {Pawlak}}, \bibinfo {author} {\bibfnamefont
  {L.}~\bibnamefont {Peng}}, \bibinfo {author} {\bibfnamefont {A.}~\bibnamefont {Ryou}}, \bibinfo {author} {\bibfnamefont {A.}~\bibnamefont {Smull}}, \bibinfo {author} {\bibfnamefont {D.}~\bibnamefont {Stack}}, \bibinfo {author} {\bibfnamefont {M.}~\bibnamefont {Stone}}, \bibinfo {author} {\bibfnamefont {A.}~\bibnamefont {Sucich}}, \bibinfo {author} {\bibfnamefont {M.}~\bibnamefont {Urbanek}}, \bibinfo {author} {\bibfnamefont {R.}~\bibnamefont {van~de Veerdonk}}, \bibinfo {author} {\bibfnamefont {Z.}~\bibnamefont {Vendeiro}}, \bibinfo {author} {\bibfnamefont {T.}~\bibnamefont {Wilkason}}, \bibinfo {author} {\bibfnamefont {T.-Y.}\ \bibnamefont {Wu}}, \bibinfo {author} {\bibfnamefont {X.}~\bibnamefont {Xie}}, \bibinfo {author} {\bibfnamefont {X.}~\bibnamefont {Zhang}},\ and\ \bibinfo {author} {\bibfnamefont {B.~J.}\ \bibnamefont {Bloom}},\ }\bibfield  {title} {\bibinfo {title} {Mid-circuit qubit measurement and rearrangement in a $^{171}${Y}b atomic array},\ }\href@noop {} {\bibfield  {journal} {\bibinfo
  {journal} {Phys. Rev. X}\ }\textbf {\bibinfo {volume} {13}},\ \bibinfo {pages} {041034} (\bibinfo {year} {2023})}\BibitemShut {NoStop}%
\bibitem [{\citenamefont {Lis}\ \emph {et~al.}(2023)\citenamefont {Lis}, \citenamefont {Senoo}, \citenamefont {McGrew}, \citenamefont {R{\"o}nchen}, \citenamefont {Jenkins},\ and\ \citenamefont {Kaufman}}]{Lis2023}%
  \BibitemOpen
  \bibfield  {author} {\bibinfo {author} {\bibfnamefont {J.~W.}\ \bibnamefont {Lis}}, \bibinfo {author} {\bibfnamefont {A.}~\bibnamefont {Senoo}}, \bibinfo {author} {\bibfnamefont {W.~F.}\ \bibnamefont {McGrew}}, \bibinfo {author} {\bibfnamefont {F.}~\bibnamefont {R{\"o}nchen}}, \bibinfo {author} {\bibfnamefont {A.}~\bibnamefont {Jenkins}},\ and\ \bibinfo {author} {\bibfnamefont {A.~M.}\ \bibnamefont {Kaufman}},\ }\bibfield  {title} {\bibinfo {title} {Mid-circuit operations using the omg-architecture in neutral atom arrays},\ }\href@noop {} {\bibfield  {journal} {\bibinfo  {journal} {Phys. Rev. X}\ }\textbf {\bibinfo {volume} {13}},\ \bibinfo {pages} {041035} (\bibinfo {year} {2023})}\BibitemShut {NoStop}%
\bibitem [{\citenamefont {Singh}\ \emph {et~al.}(2023)\citenamefont {Singh}, \citenamefont {Bradley}, \citenamefont {Anand}, \citenamefont {Ramesh}, \citenamefont {White},\ and\ \citenamefont {Bernien}}]{Singh2023}%
  \BibitemOpen
  \bibfield  {author} {\bibinfo {author} {\bibfnamefont {K.}~\bibnamefont {Singh}}, \bibinfo {author} {\bibfnamefont {C.~E.}\ \bibnamefont {Bradley}}, \bibinfo {author} {\bibfnamefont {S.}~\bibnamefont {Anand}}, \bibinfo {author} {\bibfnamefont {V.}~\bibnamefont {Ramesh}}, \bibinfo {author} {\bibfnamefont {R.}~\bibnamefont {White}},\ and\ \bibinfo {author} {\bibfnamefont {H.}~\bibnamefont {Bernien}},\ }\bibfield  {title} {\bibinfo {title} {Mid-circuit correction of correlated phase errors using an array of spectator qubits},\ }\href@noop {} {\bibfield  {journal} {\bibinfo  {journal} {Science}\ }\textbf {\bibinfo {volume} {380}},\ \bibinfo {pages} {1265} (\bibinfo {year} {2023})}\BibitemShut {NoStop}%
\bibitem [{\citenamefont {Saffman}\ and\ \citenamefont {Walker}(2005{\natexlab{a}})}]{Saffman2005a}%
  \BibitemOpen
  \bibfield  {author} {\bibinfo {author} {\bibfnamefont {M.}~\bibnamefont {Saffman}}\ and\ \bibinfo {author} {\bibfnamefont {T.~G.}\ \bibnamefont {Walker}},\ }\bibfield  {title} {\bibinfo {title} {Analysis of a quantum logic device based on dipole-dipole interactions of optically trapped {R}ydberg atoms},\ }\href@noop {} {\bibfield  {journal} {\bibinfo  {journal} {Phys. Rev. A}\ }\textbf {\bibinfo {volume} {72}},\ \bibinfo {pages} {022347} (\bibinfo {year} {2005}{\natexlab{a}})}\BibitemShut {NoStop}%
\bibitem [{\citenamefont {Saffman}\ and\ \citenamefont {Walker}(2005{\natexlab{b}})}]{Saffman2005b}%
  \BibitemOpen
  \bibfield  {author} {\bibinfo {author} {\bibfnamefont {M.}~\bibnamefont {Saffman}}\ and\ \bibinfo {author} {\bibfnamefont {T.~G.}\ \bibnamefont {Walker}},\ }\bibfield  {title} {\bibinfo {title} {Entangling single- and {N}-atom qubits for fast quantum state detection and transmission},\ }\href@noop {} {\bibfield  {journal} {\bibinfo  {journal} {Phys. Rev. A}\ }\textbf {\bibinfo {volume} {72}},\ \bibinfo {pages} {042302} (\bibinfo {year} {2005}{\natexlab{b}})}\BibitemShut {NoStop}%
\bibitem [{\citenamefont {Xu}\ \emph {et~al.}(2021)\citenamefont {Xu}, \citenamefont {Venkatramani}, \citenamefont {Cant\'u}, \citenamefont {\ifmmode~\check{S}\else \v{S}\fi{}umarac}, \citenamefont {Kl\"usener}, \citenamefont {Lukin},\ and\ \citenamefont {Vuleti\ifmmode~\acute{c}\else \'{c}\fi{}}}]{WXu2021}%
  \BibitemOpen
  \bibfield  {author} {\bibinfo {author} {\bibfnamefont {W.}~\bibnamefont {Xu}}, \bibinfo {author} {\bibfnamefont {A.~V.}\ \bibnamefont {Venkatramani}}, \bibinfo {author} {\bibfnamefont {S.~H.}\ \bibnamefont {Cant\'u}}, \bibinfo {author} {\bibfnamefont {T.}~\bibnamefont {\ifmmode~\check{S}\else \v{S}\fi{}umarac}}, \bibinfo {author} {\bibfnamefont {V.}~\bibnamefont {Kl\"usener}}, \bibinfo {author} {\bibfnamefont {M.~D.}\ \bibnamefont {Lukin}},\ and\ \bibinfo {author} {\bibfnamefont {V.}~\bibnamefont {Vuleti\ifmmode~\acute{c}\else \'{c}\fi{}}},\ }\bibfield  {title} {\bibinfo {title} {Fast preparation and detection of a {R}ydberg qubit using atomic ensembles},\ }\href {https://doi.org/10.1103/PhysRevLett.127.050501} {\bibfield  {journal} {\bibinfo  {journal} {Phys. Rev. Lett.}\ }\textbf {\bibinfo {volume} {127}},\ \bibinfo {pages} {050501} (\bibinfo {year} {2021})}\BibitemShut {NoStop}%
\bibitem [{\citenamefont {Fowler}\ \emph {et~al.}(2012)\citenamefont {Fowler}, \citenamefont {Mariantoni}, \citenamefont {Martinis},\ and\ \citenamefont {Cleland}}]{Fowler2012}%
  \BibitemOpen
  \bibfield  {author} {\bibinfo {author} {\bibfnamefont {A.~G.}\ \bibnamefont {Fowler}}, \bibinfo {author} {\bibfnamefont {M.}~\bibnamefont {Mariantoni}}, \bibinfo {author} {\bibfnamefont {J.~M.}\ \bibnamefont {Martinis}},\ and\ \bibinfo {author} {\bibfnamefont {A.~N.}\ \bibnamefont {Cleland}},\ }\bibfield  {title} {\bibinfo {title} {Surface codes: Towards practical large-scale quantum computation},\ }\href@noop {} {\bibfield  {journal} {\bibinfo  {journal} {Phys. Rev. A}\ }\textbf {\bibinfo {volume} {86}},\ \bibinfo {pages} {032324} (\bibinfo {year} {2012})}\BibitemShut {NoStop}%
\bibitem [{\citenamefont {Barredo}\ \emph {et~al.}(2016)\citenamefont {Barredo}, \citenamefont {de~Les\'el\'euc}, \citenamefont {Lienhard}, \citenamefont {Lahaye},\ and\ \citenamefont {Browaeys}}]{Barredo2016}%
  \BibitemOpen
  \bibfield  {author} {\bibinfo {author} {\bibfnamefont {D.}~\bibnamefont {Barredo}}, \bibinfo {author} {\bibfnamefont {S.}~\bibnamefont {de~Les\'el\'euc}}, \bibinfo {author} {\bibfnamefont {V.}~\bibnamefont {Lienhard}}, \bibinfo {author} {\bibfnamefont {T.}~\bibnamefont {Lahaye}},\ and\ \bibinfo {author} {\bibfnamefont {A.}~\bibnamefont {Browaeys}},\ }\bibfield  {title} {\bibinfo {title} {An atom-by-atom assembler of defect-free arbitrary two-dimensional atomic arrays},\ }\href@noop {} {\bibfield  {journal} {\bibinfo  {journal} {Science}\ }\textbf {\bibinfo {volume} {354}},\ \bibinfo {pages} {1021} (\bibinfo {year} {2016})}\BibitemShut {NoStop}%
\bibitem [{\citenamefont {Kim}\ \emph {et~al.}(2016)\citenamefont {Kim}, \citenamefont {Lee}, \citenamefont {g.~Lee}, \citenamefont {Jo}, \citenamefont {Song},\ and\ \citenamefont {Ahn}}]{Kim2016}%
  \BibitemOpen
  \bibfield  {author} {\bibinfo {author} {\bibfnamefont {H.}~\bibnamefont {Kim}}, \bibinfo {author} {\bibfnamefont {W.}~\bibnamefont {Lee}}, \bibinfo {author} {\bibfnamefont {H.}~\bibnamefont {g.~Lee}}, \bibinfo {author} {\bibfnamefont {H.}~\bibnamefont {Jo}}, \bibinfo {author} {\bibfnamefont {Y.}~\bibnamefont {Song}},\ and\ \bibinfo {author} {\bibfnamefont {J.}~\bibnamefont {Ahn}},\ }\bibfield  {title} {\bibinfo {title} {In situ single-atom array synthesis using dynamic holographic optical tweezers},\ }\href@noop {} {\bibfield  {journal} {\bibinfo  {journal} {Nat. Commun.}\ }\textbf {\bibinfo {volume} {7}},\ \bibinfo {pages} {13317} (\bibinfo {year} {2016})}\BibitemShut {NoStop}%
\bibitem [{\citenamefont {Kim}\ \emph {et~al.}(2019)\citenamefont {Kim}, \citenamefont {Keesling}, \citenamefont {Omran}, \citenamefont {Levine}, \citenamefont {Bernien}, \citenamefont {Greiner}, \citenamefont {Lukin},\ and\ \citenamefont {Englund}}]{DKim2019}%
  \BibitemOpen
  \bibfield  {author} {\bibinfo {author} {\bibfnamefont {D.}~\bibnamefont {Kim}}, \bibinfo {author} {\bibfnamefont {A.}~\bibnamefont {Keesling}}, \bibinfo {author} {\bibfnamefont {A.}~\bibnamefont {Omran}}, \bibinfo {author} {\bibfnamefont {H.}~\bibnamefont {Levine}}, \bibinfo {author} {\bibfnamefont {H.}~\bibnamefont {Bernien}}, \bibinfo {author} {\bibfnamefont {M.}~\bibnamefont {Greiner}}, \bibinfo {author} {\bibfnamefont {M.~D.}\ \bibnamefont {Lukin}},\ and\ \bibinfo {author} {\bibfnamefont {D.~R.}\ \bibnamefont {Englund}},\ }\bibfield  {title} {\bibinfo {title} {Large-scale uniform optical focus array generation with a phase spatial light modulator},\ }\href {https://doi.org/10.1364/OL.44.003178} {\bibfield  {journal} {\bibinfo  {journal} {Opt. Lett.}\ }\textbf {\bibinfo {volume} {44}},\ \bibinfo {pages} {3178} (\bibinfo {year} {2019})}\BibitemShut {NoStop}%
\bibitem [{\citenamefont {Huft}\ \emph {et~al.}(2022)\citenamefont {Huft}, \citenamefont {Song}, \citenamefont {Graham}, \citenamefont {Jooya}, \citenamefont {Deshpande}, \citenamefont {Fang}, \citenamefont {Kats},\ and\ \citenamefont {Saffman}}]{Huft2022}%
  \BibitemOpen
  \bibfield  {author} {\bibinfo {author} {\bibfnamefont {P.}~\bibnamefont {Huft}}, \bibinfo {author} {\bibfnamefont {Y.}~\bibnamefont {Song}}, \bibinfo {author} {\bibfnamefont {T.~M.}\ \bibnamefont {Graham}}, \bibinfo {author} {\bibfnamefont {K.}~\bibnamefont {Jooya}}, \bibinfo {author} {\bibfnamefont {S.}~\bibnamefont {Deshpande}}, \bibinfo {author} {\bibfnamefont {C.}~\bibnamefont {Fang}}, \bibinfo {author} {\bibfnamefont {M.}~\bibnamefont {Kats}},\ and\ \bibinfo {author} {\bibfnamefont {M.}~\bibnamefont {Saffman}},\ }\bibfield  {title} {\bibinfo {title} {Simple, passive design for large optical trap arrays for single atoms},\ }\href@noop {} {\bibfield  {journal} {\bibinfo  {journal} {Phys. Rev. A}\ }\textbf {\bibinfo {volume} {105}},\ \bibinfo {pages} {063111} (\bibinfo {year} {2022})}\BibitemShut {NoStop}%
\bibitem [{\citenamefont {Beterov}\ and\ \citenamefont {Saffman}(2015)}]{Beterov2015}%
  \BibitemOpen
  \bibfield  {author} {\bibinfo {author} {\bibfnamefont {I.~I.}\ \bibnamefont {Beterov}}\ and\ \bibinfo {author} {\bibfnamefont {M.}~\bibnamefont {Saffman}},\ }\bibfield  {title} {\bibinfo {title} {{R}ydberg blockade, {F}\"orster resonances, and quantum state measurements with different atomic species},\ }\href@noop {} {\bibfield  {journal} {\bibinfo  {journal} {Phys. Rev. A}\ }\textbf {\bibinfo {volume} {92}},\ \bibinfo {pages} {042710} (\bibinfo {year} {2015})}\BibitemShut {NoStop}%
\bibitem [{\citenamefont {Saffman}\ and\ \citenamefont {M\o{}lmer}(2009)}]{Saffman2009b}%
  \BibitemOpen
  \bibfield  {author} {\bibinfo {author} {\bibfnamefont {M.}~\bibnamefont {Saffman}}\ and\ \bibinfo {author} {\bibfnamefont {K.}~\bibnamefont {M\o{}lmer}},\ }\bibfield  {title} {\bibinfo {title} {Efficient multiparticle entanglement via asymmetric {R}ydberg blockade},\ }\href@noop {} {\bibfield  {journal} {\bibinfo  {journal} {Phys. Rev. Lett.}\ }\textbf {\bibinfo {volume} {102}},\ \bibinfo {pages} {240502} (\bibinfo {year} {2009})}\BibitemShut {NoStop}%
\bibitem [{\citenamefont {Jau}\ \emph {et~al.}(2016)\citenamefont {Jau}, \citenamefont {Hankin}, \citenamefont {Keating}, \citenamefont {Deutsch},\ and\ \citenamefont {Biedermann}}]{Jau2016}%
  \BibitemOpen
  \bibfield  {author} {\bibinfo {author} {\bibfnamefont {Y.-Y.}\ \bibnamefont {Jau}}, \bibinfo {author} {\bibfnamefont {A.~M.}\ \bibnamefont {Hankin}}, \bibinfo {author} {\bibfnamefont {T.}~\bibnamefont {Keating}}, \bibinfo {author} {\bibfnamefont {I.~H.}\ \bibnamefont {Deutsch}},\ and\ \bibinfo {author} {\bibfnamefont {G.~W.}\ \bibnamefont {Biedermann}},\ }\bibfield  {title} {\bibinfo {title} {Entangling atomic spins with a {R}ydberg-dressed spin-flip blockade},\ }\href@noop {} {\bibfield  {journal} {\bibinfo  {journal} {Nat. Phys.}\ }\textbf {\bibinfo {volume} {12}},\ \bibinfo {pages} {71} (\bibinfo {year} {2016})}\BibitemShut {NoStop}%
\bibitem [{\citenamefont {Kwon}\ \emph {et~al.}(2017)\citenamefont {Kwon}, \citenamefont {Ebert}, \citenamefont {Walker},\ and\ \citenamefont {Saffman}}]{Kwon2017}%
  \BibitemOpen
  \bibfield  {author} {\bibinfo {author} {\bibfnamefont {M.}~\bibnamefont {Kwon}}, \bibinfo {author} {\bibfnamefont {M.~F.}\ \bibnamefont {Ebert}}, \bibinfo {author} {\bibfnamefont {T.~G.}\ \bibnamefont {Walker}},\ and\ \bibinfo {author} {\bibfnamefont {M.}~\bibnamefont {Saffman}},\ }\bibfield  {title} {\bibinfo {title} {Parallel low-loss measurement of multiple atomic qubits},\ }\href@noop {} {\bibfield  {journal} {\bibinfo  {journal} {Phys. Rev. Lett.}\ }\textbf {\bibinfo {volume} {119}},\ \bibinfo {pages} {180504} (\bibinfo {year} {2017})}\BibitemShut {NoStop}%
\bibitem [{\citenamefont {Petrosyan}\ \emph {et~al.}(2017)\citenamefont {Petrosyan}, \citenamefont {Motzoi}, \citenamefont {Saffman},\ and\ \citenamefont {M\o{}lmer}}]{Petrosyan2017}%
  \BibitemOpen
  \bibfield  {author} {\bibinfo {author} {\bibfnamefont {D.}~\bibnamefont {Petrosyan}}, \bibinfo {author} {\bibfnamefont {F.}~\bibnamefont {Motzoi}}, \bibinfo {author} {\bibfnamefont {M.}~\bibnamefont {Saffman}},\ and\ \bibinfo {author} {\bibfnamefont {K.}~\bibnamefont {M\o{}lmer}},\ }\bibfield  {title} {\bibinfo {title} {High-fidelity {R}ydberg quantum gate via a two-atom dark state},\ }\href@noop {} {\bibfield  {journal} {\bibinfo  {journal} {Phys. Rev. A}\ }\textbf {\bibinfo {volume} {96}},\ \bibinfo {pages} {042306} (\bibinfo {year} {2017})}\BibitemShut {NoStop}%
\bibitem [{\citenamefont {Pillet}\ \emph {et~al.}(1984)\citenamefont {Pillet}, \citenamefont {van Linden van~den Heuvell}, \citenamefont {Smith}, \citenamefont {Kachru}, \citenamefont {Tran},\ and\ \citenamefont {Gallagher}}]{Pillet1984}%
  \BibitemOpen
  \bibfield  {author} {\bibinfo {author} {\bibfnamefont {P.}~\bibnamefont {Pillet}}, \bibinfo {author} {\bibfnamefont {H.~B.}\ \bibnamefont {van Linden van~den Heuvell}}, \bibinfo {author} {\bibfnamefont {W.~W.}\ \bibnamefont {Smith}}, \bibinfo {author} {\bibfnamefont {R.}~\bibnamefont {Kachru}}, \bibinfo {author} {\bibfnamefont {N.~H.}\ \bibnamefont {Tran}},\ and\ \bibinfo {author} {\bibfnamefont {T.~F.}\ \bibnamefont {Gallagher}},\ }\bibfield  {title} {\bibinfo {title} {Microwave ionization of {N}a {R}ydberg atoms},\ }\href {https://doi.org/10.1103/PhysRevA.30.280} {\bibfield  {journal} {\bibinfo  {journal} {Phys. Rev. A}\ }\textbf {\bibinfo {volume} {30}},\ \bibinfo {pages} {280} (\bibinfo {year} {1984})}\BibitemShut {NoStop}%
\bibitem [{\citenamefont {Fu}\ \emph {et~al.}(1990)\citenamefont {Fu}, \citenamefont {Scholz}, \citenamefont {Hettema},\ and\ \citenamefont {Gallagher}}]{PFu1990}%
  \BibitemOpen
  \bibfield  {author} {\bibinfo {author} {\bibfnamefont {P.}~\bibnamefont {Fu}}, \bibinfo {author} {\bibfnamefont {T.~J.}\ \bibnamefont {Scholz}}, \bibinfo {author} {\bibfnamefont {J.~M.}\ \bibnamefont {Hettema}},\ and\ \bibinfo {author} {\bibfnamefont {T.~F.}\ \bibnamefont {Gallagher}},\ }\bibfield  {title} {\bibinfo {title} {Ionization of {R}ydberg atoms by a circularly polarized microwave field},\ }\href {https://doi.org/10.1103/PhysRevLett.64.511} {\bibfield  {journal} {\bibinfo  {journal} {Phys. Rev. Lett.}\ }\textbf {\bibinfo {volume} {64}},\ \bibinfo {pages} {511} (\bibinfo {year} {1990})}\BibitemShut {NoStop}%
\bibitem [{\citenamefont {Knoernschild}\ \emph {et~al.}(2010)\citenamefont {Knoernschild}, \citenamefont {Zhang}, \citenamefont {Isenhower}, \citenamefont {Gill}, \citenamefont {Lu}, \citenamefont {Saffman},\ and\ \citenamefont {Kim}}]{Knoernschild2010}%
  \BibitemOpen
  \bibfield  {author} {\bibinfo {author} {\bibfnamefont {C.}~\bibnamefont {Knoernschild}}, \bibinfo {author} {\bibfnamefont {X.~L.}\ \bibnamefont {Zhang}}, \bibinfo {author} {\bibfnamefont {L.}~\bibnamefont {Isenhower}}, \bibinfo {author} {\bibfnamefont {A.~T.}\ \bibnamefont {Gill}}, \bibinfo {author} {\bibfnamefont {F.~P.}\ \bibnamefont {Lu}}, \bibinfo {author} {\bibfnamefont {M.}~\bibnamefont {Saffman}},\ and\ \bibinfo {author} {\bibfnamefont {J.}~\bibnamefont {Kim}},\ }\bibfield  {title} {\bibinfo {title} {Independent individual addressing of multiple neutral atom qubits with a {MEMS} beam steering system},\ }\href@noop {} {\bibfield  {journal} {\bibinfo  {journal} {Appl. Phys. Lett.}\ }\textbf {\bibinfo {volume} {97}},\ \bibinfo {pages} {134101} (\bibinfo {year} {2010})}\BibitemShut {NoStop}%
\bibitem [{\citenamefont {Levine}\ \emph {et~al.}(2022)\citenamefont {Levine}, \citenamefont {Bluvstein}, \citenamefont {Keesling}, \citenamefont {Wang}, \citenamefont {Ebadi}, \citenamefont {Semeghini}, \citenamefont {Omran}, \citenamefont {Greiner}, \citenamefont {Vuleti\ifmmode~\acute{c}\else \'{c}\fi{}},\ and\ \citenamefont {Lukin}}]{Levine2022}%
  \BibitemOpen
  \bibfield  {author} {\bibinfo {author} {\bibfnamefont {H.}~\bibnamefont {Levine}}, \bibinfo {author} {\bibfnamefont {D.}~\bibnamefont {Bluvstein}}, \bibinfo {author} {\bibfnamefont {A.}~\bibnamefont {Keesling}}, \bibinfo {author} {\bibfnamefont {T.~T.}\ \bibnamefont {Wang}}, \bibinfo {author} {\bibfnamefont {S.}~\bibnamefont {Ebadi}}, \bibinfo {author} {\bibfnamefont {G.}~\bibnamefont {Semeghini}}, \bibinfo {author} {\bibfnamefont {A.}~\bibnamefont {Omran}}, \bibinfo {author} {\bibfnamefont {M.}~\bibnamefont {Greiner}}, \bibinfo {author} {\bibfnamefont {V.}~\bibnamefont {Vuleti\ifmmode~\acute{c}\else \'{c}\fi{}}},\ and\ \bibinfo {author} {\bibfnamefont {M.~D.}\ \bibnamefont {Lukin}},\ }\bibfield  {title} {\bibinfo {title} {Dispersive optical systems for scalable {R}aman driving of hyperfine qubits},\ }\href {https://doi.org/10.1103/PhysRevA.105.032618} {\bibfield  {journal} {\bibinfo  {journal} {Phys. Rev. A}\ }\textbf {\bibinfo {volume} {105}},\ \bibinfo {pages} {032618} (\bibinfo {year}
  {2022})}\BibitemShut {NoStop}%
\bibitem [{\citenamefont {Phuttitarn}\ \emph {et~al.}(2023)\citenamefont {Phuttitarn}, \citenamefont {Becker}, \citenamefont {Chinnarasu}, \citenamefont {Graham},\ and\ \citenamefont {Saffman}}]{Phuttitarn2023}%
  \BibitemOpen
  \bibfield  {author} {\bibinfo {author} {\bibfnamefont {L.}~\bibnamefont {Phuttitarn}}, \bibinfo {author} {\bibfnamefont {B.~M.}\ \bibnamefont {Becker}}, \bibinfo {author} {\bibfnamefont {R.}~\bibnamefont {Chinnarasu}}, \bibinfo {author} {\bibfnamefont {T.~M.}\ \bibnamefont {Graham}},\ and\ \bibinfo {author} {\bibfnamefont {M.}~\bibnamefont {Saffman}},\ }\bibfield  {title} {\bibinfo {title} {Enhanced measurement of neutral atom qubits with machine learning},\ }\href@noop {} {\bibfield  {journal} {\bibinfo  {journal} {arXiv:2311.12217}\ } (\bibinfo {year} {2023})}\BibitemShut {NoStop}%
\bibitem [{\citenamefont {Lett}\ \emph {et~al.}(1989)\citenamefont {Lett}, \citenamefont {Phillips}, \citenamefont {Rolston}, \citenamefont {Tanner}, \citenamefont {Watts},\ and\ \citenamefont {Westbrook}}]{Lett1989}%
  \BibitemOpen
  \bibfield  {author} {\bibinfo {author} {\bibfnamefont {P.~D.}\ \bibnamefont {Lett}}, \bibinfo {author} {\bibfnamefont {W.~D.}\ \bibnamefont {Phillips}}, \bibinfo {author} {\bibfnamefont {S.~L.}\ \bibnamefont {Rolston}}, \bibinfo {author} {\bibfnamefont {C.~E.}\ \bibnamefont {Tanner}}, \bibinfo {author} {\bibfnamefont {R.~N.}\ \bibnamefont {Watts}},\ and\ \bibinfo {author} {\bibfnamefont {C.~I.}\ \bibnamefont {Westbrook}},\ }\bibfield  {title} {\bibinfo {title} {Optical molasses},\ }\href@noop {} {\bibfield  {journal} {\bibinfo  {journal} {J. Opt. Soc. Am. B}\ }\textbf {\bibinfo {volume} {6}},\ \bibinfo {pages} {2084} (\bibinfo {year} {1989})}\BibitemShut {NoStop}%
\bibitem [{\citenamefont {Su}\ \emph {et~al.}(2024)\citenamefont {Su}, \citenamefont {Douglas}, \citenamefont {Szurek}, \citenamefont {Hebert}, \citenamefont {Krahn}, \citenamefont {Groth}, \citenamefont {Phelps}, \citenamefont {Markovic},\ and\ \citenamefont {Greiner}}]{LSu2024}%
  \BibitemOpen
  \bibfield  {author} {\bibinfo {author} {\bibfnamefont {L.}~\bibnamefont {Su}}, \bibinfo {author} {\bibfnamefont {A.}~\bibnamefont {Douglas}}, \bibinfo {author} {\bibfnamefont {M.}~\bibnamefont {Szurek}}, \bibinfo {author} {\bibfnamefont {A.~H.}\ \bibnamefont {Hebert}}, \bibinfo {author} {\bibfnamefont {A.}~\bibnamefont {Krahn}}, \bibinfo {author} {\bibfnamefont {R.}~\bibnamefont {Groth}}, \bibinfo {author} {\bibfnamefont {G.~A.}\ \bibnamefont {Phelps}}, \bibinfo {author} {\bibfnamefont {O.}~\bibnamefont {Markovic}},\ and\ \bibinfo {author} {\bibfnamefont {M.}~\bibnamefont {Greiner}},\ }\bibfield  {title} {\bibinfo {title} {Fast single atom imaging in optical lattice arrays},\ }\href@noop {} {\bibfield  {journal} {\bibinfo  {journal} {arXiv:2404.09978}\ } (\bibinfo {year} {2024})}\BibitemShut {NoStop}%
\bibitem [{\citenamefont {Versluis}\ \emph {et~al.}(2017)\citenamefont {Versluis}, \citenamefont {Poletto}, \citenamefont {Khammassi}, \citenamefont {Tarasinski}, \citenamefont {Haider}, \citenamefont {Michalak}, \citenamefont {Bruno}, \citenamefont {Bertels},\ and\ \citenamefont {DiCarlo}}]{Versluis2017}%
  \BibitemOpen
  \bibfield  {author} {\bibinfo {author} {\bibfnamefont {R.}~\bibnamefont {Versluis}}, \bibinfo {author} {\bibfnamefont {S.}~\bibnamefont {Poletto}}, \bibinfo {author} {\bibfnamefont {N.}~\bibnamefont {Khammassi}}, \bibinfo {author} {\bibfnamefont {B.}~\bibnamefont {Tarasinski}}, \bibinfo {author} {\bibfnamefont {N.}~\bibnamefont {Haider}}, \bibinfo {author} {\bibfnamefont {D.~J.}\ \bibnamefont {Michalak}}, \bibinfo {author} {\bibfnamefont {A.}~\bibnamefont {Bruno}}, \bibinfo {author} {\bibfnamefont {K.}~\bibnamefont {Bertels}},\ and\ \bibinfo {author} {\bibfnamefont {L.}~\bibnamefont {DiCarlo}},\ }\bibfield  {title} {\bibinfo {title} {Scalable quantum circuit and control for a superconducting surface code},\ }\href@noop {} {\bibfield  {journal} {\bibinfo  {journal} {Phys. Rev. Appl.}\ }\textbf {\bibinfo {volume} {8}},\ \bibinfo {pages} {034021} (\bibinfo {year} {2017})}\BibitemShut {NoStop}%
\bibitem [{\citenamefont {Graham}\ \emph {et~al.}(2023{\natexlab{b}})\citenamefont {Graham}, \citenamefont {Oh},\ and\ \citenamefont {Saffman}}]{Graham2023}%
  \BibitemOpen
  \bibfield  {author} {\bibinfo {author} {\bibfnamefont {T.~M.}\ \bibnamefont {Graham}}, \bibinfo {author} {\bibfnamefont {E.}~\bibnamefont {Oh}},\ and\ \bibinfo {author} {\bibfnamefont {M.}~\bibnamefont {Saffman}},\ }\bibfield  {title} {\bibinfo {title} {Multi-scale architecture for fast optical addressing and control of large scale qubit arrays},\ }\href@noop {} {\bibfield  {journal} {\bibinfo  {journal} {Appl. Opt.}\ }\textbf {\bibinfo {volume} {62}},\ \bibinfo {pages} {3242} (\bibinfo {year} {2023}{\natexlab{b}})}\BibitemShut {NoStop}%
\bibitem [{\citenamefont {Menssen}\ \emph {et~al.}(2023)\citenamefont {Menssen}, \citenamefont {Hermans}, \citenamefont {Christen}, \citenamefont {Propson}, \citenamefont {Li}, \citenamefont {Leenheer}, \citenamefont {Zimmermann}, \citenamefont {Dong}, \citenamefont {Larocque}, \citenamefont {Raniwala}, \citenamefont {Gilbert}, \citenamefont {Eichenfield},\ and\ \citenamefont {Englund}}]{Menssen2023}%
  \BibitemOpen
  \bibfield  {author} {\bibinfo {author} {\bibfnamefont {A.~J.}\ \bibnamefont {Menssen}}, \bibinfo {author} {\bibfnamefont {A.}~\bibnamefont {Hermans}}, \bibinfo {author} {\bibfnamefont {I.}~\bibnamefont {Christen}}, \bibinfo {author} {\bibfnamefont {T.}~\bibnamefont {Propson}}, \bibinfo {author} {\bibfnamefont {C.}~\bibnamefont {Li}}, \bibinfo {author} {\bibfnamefont {A.~J.}\ \bibnamefont {Leenheer}}, \bibinfo {author} {\bibfnamefont {M.}~\bibnamefont {Zimmermann}}, \bibinfo {author} {\bibfnamefont {M.}~\bibnamefont {Dong}}, \bibinfo {author} {\bibfnamefont {H.}~\bibnamefont {Larocque}}, \bibinfo {author} {\bibfnamefont {H.}~\bibnamefont {Raniwala}}, \bibinfo {author} {\bibfnamefont {G.}~\bibnamefont {Gilbert}}, \bibinfo {author} {\bibfnamefont {M.}~\bibnamefont {Eichenfield}},\ and\ \bibinfo {author} {\bibfnamefont {D.~R.}\ \bibnamefont {Englund}},\ }\bibfield  {title} {\bibinfo {title} {Scalable photonic integrated circuits for high-fidelity light control},\ }\href@noop {} {\bibfield  {journal} {\bibinfo
  {journal} {Optica}\ }\textbf {\bibinfo {volume} {10}},\ \bibinfo {pages} {1366} (\bibinfo {year} {2023})}\BibitemShut {NoStop}%
\bibitem [{\citenamefont {Zhang}\ \emph {et~al.}(2024)\citenamefont {Zhang}, \citenamefont {Peng}, \citenamefont {Paul},\ and\ \citenamefont {Thompson}}]{BZhang2024}%
  \BibitemOpen
  \bibfield  {author} {\bibinfo {author} {\bibfnamefont {B.}~\bibnamefont {Zhang}}, \bibinfo {author} {\bibfnamefont {P.}~\bibnamefont {Peng}}, \bibinfo {author} {\bibfnamefont {A.}~\bibnamefont {Paul}},\ and\ \bibinfo {author} {\bibfnamefont {J.~D.}\ \bibnamefont {Thompson}},\ }\bibfield  {title} {\bibinfo {title} {Scaled local gate controller for optically addressed qubits},\ }\href {https://doi.org/10.1364/OPTICA.512155} {\bibfield  {journal} {\bibinfo  {journal} {Optica}\ }\textbf {\bibinfo {volume} {11}},\ \bibinfo {pages} {227} (\bibinfo {year} {2024})}\BibitemShut {NoStop}%
\bibitem [{\citenamefont {Ulku}\ \emph {et~al.}(2019)\citenamefont {Ulku}, \citenamefont {Bruschini}, \citenamefont {Antolovi\'c}, \citenamefont {Kuo}, \citenamefont {Ankri}, \citenamefont {Weiss}, \citenamefont {Michalet},\ and\ \citenamefont {Charbon}}]{Ulku2019}%
  \BibitemOpen
  \bibfield  {author} {\bibinfo {author} {\bibfnamefont {A.~C.}\ \bibnamefont {Ulku}}, \bibinfo {author} {\bibfnamefont {C.}~\bibnamefont {Bruschini}}, \bibinfo {author} {\bibfnamefont {I.~M.}\ \bibnamefont {Antolovi\'c}}, \bibinfo {author} {\bibfnamefont {Y.}~\bibnamefont {Kuo}}, \bibinfo {author} {\bibfnamefont {R.}~\bibnamefont {Ankri}}, \bibinfo {author} {\bibfnamefont {S.}~\bibnamefont {Weiss}}, \bibinfo {author} {\bibfnamefont {X.}~\bibnamefont {Michalet}},\ and\ \bibinfo {author} {\bibfnamefont {E.}~\bibnamefont {Charbon}},\ }\bibfield  {title} {\bibinfo {title} {A 512 × 512 {SPAD} image sensor with integrated gating for widefield {FLIM}},\ }\href@noop {} {\bibfield  {journal} {\bibinfo  {journal} {IEEE J. Sel. Top. Quant. Electr.}\ }\textbf {\bibinfo {volume} {25}},\ \bibinfo {pages} {6801212} (\bibinfo {year} {2019})}\BibitemShut {NoStop}%
\bibitem [{\citenamefont {Brion}\ \emph {et~al.}(2007)\citenamefont {Brion}, \citenamefont {Mouritzen},\ and\ \citenamefont {M\o{}lmer}}]{Brion2007c}%
  \BibitemOpen
  \bibfield  {author} {\bibinfo {author} {\bibfnamefont {E.}~\bibnamefont {Brion}}, \bibinfo {author} {\bibfnamefont {A.~S.}\ \bibnamefont {Mouritzen}},\ and\ \bibinfo {author} {\bibfnamefont {K.}~\bibnamefont {M\o{}lmer}},\ }\bibfield  {title} {\bibinfo {title} {Conditional dynamics induced by new configurations for {R}ydberg dipole-dipole interactions},\ }\href@noop {} {\bibfield  {journal} {\bibinfo  {journal} {Phys. Rev. A}\ }\textbf {\bibinfo {volume} {76}},\ \bibinfo {pages} {022334} (\bibinfo {year} {2007})}\BibitemShut {NoStop}%
\bibitem [{\citenamefont {Isenhower}\ \emph {et~al.}(2011)\citenamefont {Isenhower}, \citenamefont {Saffman},\ and\ \citenamefont {M\o{}lmer}}]{Isenhower2011}%
  \BibitemOpen
  \bibfield  {author} {\bibinfo {author} {\bibfnamefont {L.}~\bibnamefont {Isenhower}}, \bibinfo {author} {\bibfnamefont {M.}~\bibnamefont {Saffman}},\ and\ \bibinfo {author} {\bibfnamefont {K.}~\bibnamefont {M\o{}lmer}},\ }\bibfield  {title} {\bibinfo {title} {Multibit {C}$_k${NOT} quantum gates via {R}ydberg blockade},\ }\href@noop {} {\bibfield  {journal} {\bibinfo  {journal} {Quant. Inf. Proc.}\ }\textbf {\bibinfo {volume} {10}},\ \bibinfo {pages} {755} (\bibinfo {year} {2011})}\BibitemShut {NoStop}%
\bibitem [{\citenamefont {Petrosyan}\ \emph {et~al.}(2016)\citenamefont {Petrosyan}, \citenamefont {Saffman},\ and\ \citenamefont {M\o{}lmer}}]{Petrosyan2016}%
  \BibitemOpen
  \bibfield  {author} {\bibinfo {author} {\bibfnamefont {D.}~\bibnamefont {Petrosyan}}, \bibinfo {author} {\bibfnamefont {M.}~\bibnamefont {Saffman}},\ and\ \bibinfo {author} {\bibfnamefont {K.}~\bibnamefont {M\o{}lmer}},\ }\bibfield  {title} {\bibinfo {title} {Grover search algorithm with {R}ydberg-blockaded atoms: quantum {M}onte-{C}arlo simulations},\ }\href@noop {} {\bibfield  {journal} {\bibinfo  {journal} {J. Phys. B: At. Mol. Opt. Phys.}\ }\textbf {\bibinfo {volume} {49}},\ \bibinfo {pages} {094004} (\bibinfo {year} {2016})}\BibitemShut {NoStop}%
\bibitem [{\citenamefont {Su}\ \emph {et~al.}(2018)\citenamefont {Su}, \citenamefont {Shen}, \citenamefont {Erjun},\ and\ \citenamefont {Shou}}]{Su2018}%
  \BibitemOpen
  \bibfield  {author} {\bibinfo {author} {\bibfnamefont {S.}~\bibnamefont {Su}}, \bibinfo {author} {\bibfnamefont {H.}~\bibnamefont {Shen}}, \bibinfo {author} {\bibfnamefont {L.}~\bibnamefont {Erjun}},\ and\ \bibinfo {author} {\bibfnamefont {Z.}~\bibnamefont {Shou}},\ }\bibfield  {title} {\bibinfo {title} {One-step construction of the multiple-qubit {R}ydberg controlled-{PHASE} gate},\ }\href@noop {} {\bibfield  {journal} {\bibinfo  {journal} {Phys. Rev. A}\ }\textbf {\bibinfo {volume} {98}},\ \bibinfo {pages} {032306} (\bibinfo {year} {2018})}\BibitemShut {NoStop}%
\bibitem [{\citenamefont {Shi}(2018)}]{XFShi2018}%
  \BibitemOpen
  \bibfield  {author} {\bibinfo {author} {\bibfnamefont {X.-F.}\ \bibnamefont {Shi}},\ }\bibfield  {title} {\bibinfo {title} {Deutsch, {T}offoli, and {CNOT} gates via {R}ydberg blockade of neutral atoms},\ }\href@noop {} {\bibfield  {journal} {\bibinfo  {journal} {Phys. Rev. Applied}\ }\textbf {\bibinfo {volume} {9}},\ \bibinfo {pages} {051001} (\bibinfo {year} {2018})}\BibitemShut {NoStop}%
\bibitem [{\citenamefont {Beterov}\ \emph {et~al.}(2018)\citenamefont {Beterov}, \citenamefont {Ashkarin}, \citenamefont {Yakshina}, \citenamefont {Tretyakov}, \citenamefont {Entin}, \citenamefont {Ryabtsev}, \citenamefont {Cheinet}, \citenamefont {Pillet},\ and\ \citenamefont {Saffman}}]{Beterov2018b}%
  \BibitemOpen
  \bibfield  {author} {\bibinfo {author} {\bibfnamefont {I.~I.}\ \bibnamefont {Beterov}}, \bibinfo {author} {\bibfnamefont {I.~N.}\ \bibnamefont {Ashkarin}}, \bibinfo {author} {\bibfnamefont {E.~A.}\ \bibnamefont {Yakshina}}, \bibinfo {author} {\bibfnamefont {D.~B.}\ \bibnamefont {Tretyakov}}, \bibinfo {author} {\bibfnamefont {V.~M.}\ \bibnamefont {Entin}}, \bibinfo {author} {\bibfnamefont {I.~I.}\ \bibnamefont {Ryabtsev}}, \bibinfo {author} {\bibfnamefont {P.}~\bibnamefont {Cheinet}}, \bibinfo {author} {\bibfnamefont {P.}~\bibnamefont {Pillet}},\ and\ \bibinfo {author} {\bibfnamefont {M.}~\bibnamefont {Saffman}},\ }\bibfield  {title} {\bibinfo {title} {Fast three-qubit {T}offoli quantum gate based on three-body {F}\"orster resonances in {R}ydberg atoms},\ }\href@noop {} {\bibfield  {journal} {\bibinfo  {journal} {Phys. Rev. A}\ }\textbf {\bibinfo {volume} {98}},\ \bibinfo {pages} {042704} (\bibinfo {year} {2018})}\BibitemShut {NoStop}%
\bibitem [{\citenamefont {Khazali}\ and\ \citenamefont {M\o{}lmer}(2020)}]{Khazali2020}%
  \BibitemOpen
  \bibfield  {author} {\bibinfo {author} {\bibfnamefont {M.}~\bibnamefont {Khazali}}\ and\ \bibinfo {author} {\bibfnamefont {K.}~\bibnamefont {M\o{}lmer}},\ }\bibfield  {title} {\bibinfo {title} {Fast multiqubit gates by adiabatic evolution in interacting excited-state manifolds of {R}ydberg atoms and superconducting circuits},\ }\href@noop {} {\bibfield  {journal} {\bibinfo  {journal} {Phys. Rev. X}\ }\textbf {\bibinfo {volume} {10}},\ \bibinfo {pages} {021054} (\bibinfo {year} {2020})}\BibitemShut {NoStop}%
\bibitem [{\citenamefont {Young}\ \emph {et~al.}(2021)\citenamefont {Young}, \citenamefont {Bienias}, \citenamefont {Belyansky}, \citenamefont {Kaufman},\ and\ \citenamefont {Gorshkov}}]{Young2021}%
  \BibitemOpen
  \bibfield  {author} {\bibinfo {author} {\bibfnamefont {J.~T.}\ \bibnamefont {Young}}, \bibinfo {author} {\bibfnamefont {P.}~\bibnamefont {Bienias}}, \bibinfo {author} {\bibfnamefont {R.}~\bibnamefont {Belyansky}}, \bibinfo {author} {\bibfnamefont {A.~M.}\ \bibnamefont {Kaufman}},\ and\ \bibinfo {author} {\bibfnamefont {A.~V.}\ \bibnamefont {Gorshkov}},\ }\bibfield  {title} {\bibinfo {title} {Asymmetric blockade and multiqubit gates via dipole-dipole interactions},\ }\href {https://doi.org/10.1103/PhysRevLett.127.120501} {\bibfield  {journal} {\bibinfo  {journal} {Phys. Rev. Lett.}\ }\textbf {\bibinfo {volume} {127}},\ \bibinfo {pages} {120501} (\bibinfo {year} {2021})}\BibitemShut {NoStop}%
\bibitem [{\citenamefont {Rasmussen}\ \emph {et~al.}(2020)\citenamefont {Rasmussen}, \citenamefont {Groenland}, \citenamefont {Gerritsma}, \citenamefont {Schoutens},\ and\ \citenamefont {Zinner}}]{Rasmussen2020}%
  \BibitemOpen
  \bibfield  {author} {\bibinfo {author} {\bibfnamefont {S.~E.}\ \bibnamefont {Rasmussen}}, \bibinfo {author} {\bibfnamefont {K.}~\bibnamefont {Groenland}}, \bibinfo {author} {\bibfnamefont {R.}~\bibnamefont {Gerritsma}}, \bibinfo {author} {\bibfnamefont {K.}~\bibnamefont {Schoutens}},\ and\ \bibinfo {author} {\bibfnamefont {N.~T.}\ \bibnamefont {Zinner}},\ }\bibfield  {title} {\bibinfo {title} {Single-step implementation of high-fidelity $n$-bit {T}offoli gates},\ }\href@noop {} {\bibfield  {journal} {\bibinfo  {journal} {Phys. Rev. A}\ }\textbf {\bibinfo {volume} {101}},\ \bibinfo {pages} {022308} (\bibinfo {year} {2020})}\BibitemShut {NoStop}%
\bibitem [{\citenamefont {Xing}\ \emph {et~al.}(2020)\citenamefont {Xing}, \citenamefont {Wu},\ and\ \citenamefont {Xu}}]{THXing2020}%
  \BibitemOpen
  \bibfield  {author} {\bibinfo {author} {\bibfnamefont {T.~H.}\ \bibnamefont {Xing}}, \bibinfo {author} {\bibfnamefont {X.}~\bibnamefont {Wu}},\ and\ \bibinfo {author} {\bibfnamefont {G.~F.}\ \bibnamefont {Xu}},\ }\bibfield  {title} {\bibinfo {title} {Nonadiabatic holonomic three-qubit controlled gates realized by one-shot implementation},\ }\href@noop {} {\bibfield  {journal} {\bibinfo  {journal} {Phys. Rev. A}\ }\textbf {\bibinfo {volume} {101}},\ \bibinfo {pages} {012306} (\bibinfo {year} {2020})}\BibitemShut {NoStop}%
\bibitem [{\citenamefont {Wu}\ \emph {et~al.}(2021)\citenamefont {Wu}, \citenamefont {Wang}, \citenamefont {Han}, \citenamefont {Su}, \citenamefont {Xia}, \citenamefont {Jiang},\ and\ \citenamefont {Song}}]{JLWu2021}%
  \BibitemOpen
  \bibfield  {author} {\bibinfo {author} {\bibfnamefont {J.-L.}\ \bibnamefont {Wu}}, \bibinfo {author} {\bibfnamefont {Y.}~\bibnamefont {Wang}}, \bibinfo {author} {\bibfnamefont {J.-X.}\ \bibnamefont {Han}}, \bibinfo {author} {\bibfnamefont {S.-L.}\ \bibnamefont {Su}}, \bibinfo {author} {\bibfnamefont {Y.}~\bibnamefont {Xia}}, \bibinfo {author} {\bibfnamefont {Y.}~\bibnamefont {Jiang}},\ and\ \bibinfo {author} {\bibfnamefont {J.}~\bibnamefont {Song}},\ }\bibfield  {title} {\bibinfo {title} {Resilient quantum gates on periodically driven {R}ydberg atoms},\ }\href@noop {} {\bibfield  {journal} {\bibinfo  {journal} {Phys. Rev. A}\ }\textbf {\bibinfo {volume} {103}},\ \bibinfo {pages} {012601} (\bibinfo {year} {2021})}\BibitemShut {NoStop}%
\bibitem [{\citenamefont {He}\ \emph {et~al.}(2022)\citenamefont {He}, \citenamefont {Liu}, \citenamefont {Guo}, \citenamefont {Yan}, \citenamefont {Luo}, \citenamefont {Liang}, \citenamefont {Su},\ and\ \citenamefont {Feng}}]{YHe2022}%
  \BibitemOpen
  \bibfield  {author} {\bibinfo {author} {\bibfnamefont {Y.}~\bibnamefont {He}}, \bibinfo {author} {\bibfnamefont {J.-X.}\ \bibnamefont {Liu}}, \bibinfo {author} {\bibfnamefont {F.-Q.}\ \bibnamefont {Guo}}, \bibinfo {author} {\bibfnamefont {L.-L.}\ \bibnamefont {Yan}}, \bibinfo {author} {\bibfnamefont {R.}~\bibnamefont {Luo}}, \bibinfo {author} {\bibfnamefont {E.}~\bibnamefont {Liang}}, \bibinfo {author} {\bibfnamefont {S.-L.}\ \bibnamefont {Su}},\ and\ \bibinfo {author} {\bibfnamefont {M.}~\bibnamefont {Feng}},\ }\bibfield  {title} {\bibinfo {title} {Multiple-qubit {R}ydberg quantum logic gate via dressed-state scheme},\ }\href@noop {} {\bibfield  {journal} {\bibinfo  {journal} {Opt. Commun.}\ }\textbf {\bibinfo {volume} {505}},\ \bibinfo {pages} {127500} (\bibinfo {year} {2022})}\BibitemShut {NoStop}%
\bibitem [{\citenamefont {Li}\ \emph {et~al.}(2021)\citenamefont {Li}, \citenamefont {Guo}, \citenamefont {Jin}, \citenamefont {Yan}, \citenamefont {Liang},\ and\ \citenamefont {Su}}]{MLi2021}%
  \BibitemOpen
  \bibfield  {author} {\bibinfo {author} {\bibfnamefont {M.}~\bibnamefont {Li}}, \bibinfo {author} {\bibfnamefont {F.-Q.}\ \bibnamefont {Guo}}, \bibinfo {author} {\bibfnamefont {Z.}~\bibnamefont {Jin}}, \bibinfo {author} {\bibfnamefont {L.-L.}\ \bibnamefont {Yan}}, \bibinfo {author} {\bibfnamefont {E.-J.}\ \bibnamefont {Liang}},\ and\ \bibinfo {author} {\bibfnamefont {S.-L.}\ \bibnamefont {Su}},\ }\bibfield  {title} {\bibinfo {title} {Multiple-qubit controlled unitary quantum gate for {R}ydberg atoms using shortcut to adiabaticity and optimized geometric quantum operations},\ }\href@noop {} {\bibfield  {journal} {\bibinfo  {journal} {Phys. Rev. A}\ }\textbf {\bibinfo {volume} {103}},\ \bibinfo {pages} {062607} (\bibinfo {year} {2021})}\BibitemShut {NoStop}%
\bibitem [{\citenamefont {Pelegr\'i}\ \emph {et~al.}(2022)\citenamefont {Pelegr\'i}, \citenamefont {Daley},\ and\ \citenamefont {Pritchard}}]{Pelegri2022}%
  \BibitemOpen
  \bibfield  {author} {\bibinfo {author} {\bibfnamefont {G.}~\bibnamefont {Pelegr\'i}}, \bibinfo {author} {\bibfnamefont {A.~J.}\ \bibnamefont {Daley}},\ and\ \bibinfo {author} {\bibfnamefont {J.~D.}\ \bibnamefont {Pritchard}},\ }\bibfield  {title} {\bibinfo {title} {High-fidelity multiqubit {R}ydberg gates via two-photon adiabatic rapid passage},\ }\href@noop {} {\bibfield  {journal} {\bibinfo  {journal} {Qu. Sci Technol.}\ }\textbf {\bibinfo {volume} {7}},\ \bibinfo {pages} {045020} (\bibinfo {year} {2022})}\BibitemShut {NoStop}%
\bibitem [{\citenamefont {Farouk}\ \emph {et~al.}(2022)\citenamefont {Farouk}, \citenamefont {Beterov}, \citenamefont {Xu}, \citenamefont {Bergamini},\ and\ \citenamefont {Ryabtsev}}]{Farouk2022}%
  \BibitemOpen
  \bibfield  {author} {\bibinfo {author} {\bibfnamefont {A.~M.}\ \bibnamefont {Farouk}}, \bibinfo {author} {\bibfnamefont {I.}~\bibnamefont {Beterov}}, \bibinfo {author} {\bibfnamefont {P.}~\bibnamefont {Xu}}, \bibinfo {author} {\bibfnamefont {S.}~\bibnamefont {Bergamini}},\ and\ \bibinfo {author} {\bibfnamefont {I.}~\bibnamefont {Ryabtsev}},\ }\bibfield  {title} {\bibinfo {title} {Parallel implementation of {CNOT$^{\rm N}$} and {$\rm C_2NOT^2$} gates via homonuclear and heteronuclear {F}\"orster interactions of {R}ydberg atoms},\ }\href@noop {} {\bibfield  {journal} {\bibinfo  {journal} {arXiv:2206.12176}\ } (\bibinfo {year} {2022})}\BibitemShut {NoStop}%
\bibitem [{\citenamefont {Kinos}\ and\ \citenamefont {M\o{}lmer}(2023)}]{Kinos2023}%
  \BibitemOpen
  \bibfield  {author} {\bibinfo {author} {\bibfnamefont {A.}~\bibnamefont {Kinos}}\ and\ \bibinfo {author} {\bibfnamefont {K.}~\bibnamefont {M\o{}lmer}},\ }\bibfield  {title} {\bibinfo {title} {Optical multiqubit gate operations on an excitation-blockaded atomic quantum register},\ }\href {https://doi.org/10.1103/PhysRevResearch.5.013205} {\bibfield  {journal} {\bibinfo  {journal} {Phys. Rev. Res.}\ }\textbf {\bibinfo {volume} {5}},\ \bibinfo {pages} {013205} (\bibinfo {year} {2023})}\BibitemShut {NoStop}%
\bibitem [{\citenamefont {Levine}\ \emph {et~al.}(2019)\citenamefont {Levine}, \citenamefont {Keesling}, \citenamefont {Semeghini}, \citenamefont {Omran}, \citenamefont {Wang}, \citenamefont {Ebadi}, \citenamefont {Bernien}, \citenamefont {Greiner}, \citenamefont {Vuleti\'c}, \citenamefont {Pichler},\ and\ \citenamefont {Lukin}}]{Levine2019}%
  \BibitemOpen
  \bibfield  {author} {\bibinfo {author} {\bibfnamefont {H.}~\bibnamefont {Levine}}, \bibinfo {author} {\bibfnamefont {A.}~\bibnamefont {Keesling}}, \bibinfo {author} {\bibfnamefont {G.}~\bibnamefont {Semeghini}}, \bibinfo {author} {\bibfnamefont {A.}~\bibnamefont {Omran}}, \bibinfo {author} {\bibfnamefont {T.~T.}\ \bibnamefont {Wang}}, \bibinfo {author} {\bibfnamefont {S.}~\bibnamefont {Ebadi}}, \bibinfo {author} {\bibfnamefont {H.}~\bibnamefont {Bernien}}, \bibinfo {author} {\bibfnamefont {M.}~\bibnamefont {Greiner}}, \bibinfo {author} {\bibfnamefont {V.}~\bibnamefont {Vuleti\'c}}, \bibinfo {author} {\bibfnamefont {H.}~\bibnamefont {Pichler}},\ and\ \bibinfo {author} {\bibfnamefont {M.~D.}\ \bibnamefont {Lukin}},\ }\bibfield  {title} {\bibinfo {title} {Parallel implementation of high-fidelity multiqubit gates with neutral atoms},\ }\href@noop {} {\bibfield  {journal} {\bibinfo  {journal} {Phys. Rev. Lett.}\ }\textbf {\bibinfo {volume} {123}},\ \bibinfo {pages} {170503} (\bibinfo {year} {2019})}\BibitemShut
  {NoStop}%
\bibitem [{\citenamefont {Cao}\ \emph {et~al.}(2024)\citenamefont {Cao}, \citenamefont {Eckner}, \citenamefont {Yelin}, \citenamefont {Young}, \citenamefont {Jandura}, \citenamefont {Yan}, \citenamefont {Kim}, \citenamefont {Pupillo}, \citenamefont {Ye}, \citenamefont {Oppong},\ and\ \citenamefont {Kaufman}}]{ACao2024}%
  \BibitemOpen
  \bibfield  {author} {\bibinfo {author} {\bibfnamefont {A.}~\bibnamefont {Cao}}, \bibinfo {author} {\bibfnamefont {W.~J.}\ \bibnamefont {Eckner}}, \bibinfo {author} {\bibfnamefont {T.~L.}\ \bibnamefont {Yelin}}, \bibinfo {author} {\bibfnamefont {A.~W.}\ \bibnamefont {Young}}, \bibinfo {author} {\bibfnamefont {S.}~\bibnamefont {Jandura}}, \bibinfo {author} {\bibfnamefont {L.}~\bibnamefont {Yan}}, \bibinfo {author} {\bibfnamefont {K.}~\bibnamefont {Kim}}, \bibinfo {author} {\bibfnamefont {G.}~\bibnamefont {Pupillo}}, \bibinfo {author} {\bibfnamefont {J.}~\bibnamefont {Ye}}, \bibinfo {author} {\bibfnamefont {N.~D.}\ \bibnamefont {Oppong}},\ and\ \bibinfo {author} {\bibfnamefont {A.~M.}\ \bibnamefont {Kaufman}},\ }\bibfield  {title} {\bibinfo {title} {Multi-qubit gates and '{S}chr\"odinger cat' states in an optical clock},\ }\href@noop {} {\bibfield  {journal} {\bibinfo  {journal} {arXiv:2402.16289}\ } (\bibinfo {year} {2024})}\BibitemShut {NoStop}%
\bibitem [{\citenamefont {Roos}\ and\ \citenamefont {M\o{}lmer}(2004)}]{Roos2004}%
  \BibitemOpen
  \bibfield  {author} {\bibinfo {author} {\bibfnamefont {I.}~\bibnamefont {Roos}}\ and\ \bibinfo {author} {\bibfnamefont {K.}~\bibnamefont {M\o{}lmer}},\ }\bibfield  {title} {\bibinfo {title} {Quantum computing with an inhomogeneously broadened ensemble of ions: Suppression of errors from detuning variations by specially adapted pulses and coherent population trapping},\ }\href@noop {} {\bibfield  {journal} {\bibinfo  {journal} {Phys. Rev. A}\ }\textbf {\bibinfo {volume} {69}},\ \bibinfo {pages} {022321} (\bibinfo {year} {2004})}\BibitemShut {NoStop}%
\bibitem [{\citenamefont {Corlett}\ \emph {et~al.}(2024)\citenamefont {Corlett}, \citenamefont {\v{C}epait\.{e}}, \citenamefont {Daley}, \citenamefont {Gustiani}, \citenamefont {Pelegrí}, \citenamefont {Pritchard}, \citenamefont {Linden},\ and\ \citenamefont {Skrzypczyk}}]{Corlett2024}%
  \BibitemOpen
  \bibfield  {author} {\bibinfo {author} {\bibfnamefont {C.}~\bibnamefont {Corlett}}, \bibinfo {author} {\bibfnamefont {I.}~\bibnamefont {\v{C}epait\.{e}}}, \bibinfo {author} {\bibfnamefont {A.~J.}\ \bibnamefont {Daley}}, \bibinfo {author} {\bibfnamefont {C.}~\bibnamefont {Gustiani}}, \bibinfo {author} {\bibfnamefont {G.}~\bibnamefont {Pelegrí}}, \bibinfo {author} {\bibfnamefont {J.~D.}\ \bibnamefont {Pritchard}}, \bibinfo {author} {\bibfnamefont {N.}~\bibnamefont {Linden}},\ and\ \bibinfo {author} {\bibfnamefont {P.}~\bibnamefont {Skrzypczyk}},\ }\bibfield  {title} {\bibinfo {title} {Speeding up quantum measurement using space-time trade-off},\ }\href@noop {} {\bibfield  {journal} {\bibinfo  {journal} {arXiv:2407.17342}\ } (\bibinfo {year} {2024})}\BibitemShut {NoStop}%
\bibitem [{\citenamefont {Beterov}\ \emph {et~al.}(2009)\citenamefont {Beterov}, \citenamefont {Ryabtsev}, \citenamefont {Tretyakov},\ and\ \citenamefont {Entin}}]{Beterov2009}%
  \BibitemOpen
  \bibfield  {author} {\bibinfo {author} {\bibfnamefont {I.~I.}\ \bibnamefont {Beterov}}, \bibinfo {author} {\bibfnamefont {I.~I.}\ \bibnamefont {Ryabtsev}}, \bibinfo {author} {\bibfnamefont {D.~B.}\ \bibnamefont {Tretyakov}},\ and\ \bibinfo {author} {\bibfnamefont {V.~M.}\ \bibnamefont {Entin}},\ }\bibfield  {title} {\bibinfo {title} {Quasiclassical calculations of blackbody-radiation-induced depopulation rates and effective lifetimes of {R}ydberg n{S}, n{P}, and n{D} alkali-metal atoms with $n \le 80$},\ }\href@noop {} {\bibfield  {journal} {\bibinfo  {journal} {Phys. Rev. A}\ }\textbf {\bibinfo {volume} {79}},\ \bibinfo {pages} {052504} (\bibinfo {year} {2009})}\BibitemShut {NoStop}%
\bibitem [{\citenamefont {Robertson}\ \emph {et~al.}(2021)\citenamefont {Robertson}, \citenamefont {Šibalić}, \citenamefont {Potvliege},\ and\ \citenamefont {Jones}}]{ARC3.0}%
  \BibitemOpen
  \bibfield  {author} {\bibinfo {author} {\bibfnamefont {E.}~\bibnamefont {Robertson}}, \bibinfo {author} {\bibfnamefont {N.}~\bibnamefont {Šibalić}}, \bibinfo {author} {\bibfnamefont {R.}~\bibnamefont {Potvliege}},\ and\ \bibinfo {author} {\bibfnamefont {M.}~\bibnamefont {Jones}},\ }\bibfield  {title} {\bibinfo {title} {{ARC} 3.0: An expanded {P}ython toolbox for atomic physics calculations},\ }\href {https://doi.org/https://doi.org/10.1016/j.cpc.2020.107814} {\bibfield  {journal} {\bibinfo  {journal} {Comp. Phys. Commun.}\ }\textbf {\bibinfo {volume} {261}},\ \bibinfo {pages} {107814} (\bibinfo {year} {2021})}\BibitemShut {NoStop}%
\bibitem [{\citenamefont {Šibalić}\ \emph {et~al.}(2017)\citenamefont {Šibalić}, \citenamefont {Pritchard}, \citenamefont {Adams},\ and\ \citenamefont {Weatherill}}]{ARC}%
  \BibitemOpen
  \bibfield  {author} {\bibinfo {author} {\bibfnamefont {N.}~\bibnamefont {Šibalić}}, \bibinfo {author} {\bibfnamefont {J.}~\bibnamefont {Pritchard}}, \bibinfo {author} {\bibfnamefont {C.}~\bibnamefont {Adams}},\ and\ \bibinfo {author} {\bibfnamefont {K.}~\bibnamefont {Weatherill}},\ }\bibfield  {title} {\bibinfo {title} {{ARC}: An open-source library for calculating properties of alkali {R}ydberg atoms},\ }\href {https://doi.org/https://doi.org/10.1016/j.cpc.2017.06.015} {\bibfield  {journal} {\bibinfo  {journal} {Comp. Phys. Commun.}\ }\textbf {\bibinfo {volume} {220}},\ \bibinfo {pages} {319} (\bibinfo {year} {2017})}\BibitemShut {NoStop}%
\end{thebibliography}%

\end{document}